\documentclass[12pt]{iopart}
\usepackage{iopams}
\expandafter\let\csname equation*\endcsname\relax

\expandafter\let\csname endequation*\endcsname\relax

\usepackage{bm,bbm}
\usepackage{amsmath}
\usepackage{amsfonts}
\usepackage{amstext,amsthm}
\usepackage{amssymb}   
\usepackage{graphicx}
\usepackage{verbatim}
\usepackage{cite}

\begin{document}

\hyphenation{ge-ne-ra-li-zed}
\def\hf{\hat{f}}
\def\Ord{\mathcal{O}}
\newcommand{\barr}{\begin{eqnarray}}
\newcommand{\earr}{\end{eqnarray}}
\newcommand{\beq}{\begin{equation}}
\newcommand{\eeq}{\end{equation}}
\newcommand{\be}{\begin{equation}}
\newcommand{\ee}{\end{equation}}
\newcommand{\ra}{\right\rangle}
\newcommand{\de}{\mathrm{d}}
\newcommand{\la}{\left\langle}
\newcommand{\correc}{ \textcolor{red} }
\newtheorem{theorem}{Theorem}

\newcommand{\gv}[1]{\ensuremath{\mbox{\boldmath$ #1 $}}} 
\newcommand{\uv}[1]{\ensuremath{\mathbf{\hat{#1}}}} 
\newcommand{\abs}[1]{\left| #1 \right|} 
\let\underdot=\d 
\renewcommand{\d}[2]{\frac{d #1}{d #2}} 
\newcommand{\dd}[2]{\frac{d^2 #1}{d #2^2}} 
\newcommand{\pd}[2]{\frac{\partial #1}{\partial #2}} 
\newcommand{\pdd}[2]{\frac{\partial^2 #1}{\partial #2^2}} 
\newcommand{\pdc}[3]{\left( \frac{\partial #1}{\partial #2}
 \right)_{#3}} 
\newcommand{\ket}[1]{\left| #1 \right>} 
\newcommand{\bra}[1]{\left< #1 \right|} 
\newcommand{\braket}[2]{\left< #1 \vphantom{#2} \right|
 \left. #2 \vphantom{#1} \right>} 
\newcommand{\matrixel}[3]{\left< #1 \vphantom{#2#3} \right|
 #2 \left| #3 \vphantom{#1#2} \right>} 
\newcommand{\grad}[1]{\gv{\nabla} #1} 
\let\divsymb=\div 
\renewcommand{\div}[1]{\gv{\nabla} \cdot #1} 
\newcommand{\curl}[1]{\gv{\nabla} \times #1} 
\let\baraccent=\= 
\renewcommand{\=}[1]{\stackrel{#1}{=}} 

\newcommand{\numberset}{\mathbb}
\newcommand{\N}{\numberset{N}}
\newcommand{\Z}{\numberset{Z}}
\newcommand{\R}{\numberset{R}}
\newcommand{\C}{\numberset{C}}

\newcommand{\Res}{\mathrm{Res}}
\newcommand{\Var}{\mathrm{Var}}
\newcommand{\Cov}{\mathrm{Cov}}
\newcommand{\avg}[1]{\left< #1 \right>} 
\newcommand{\T}{\mathcal{T}}
\newcommand{\rs}{\rho^{\star}}
\def\im{{\rm i}}
\newcommand{\kk}{\kappa}
\newcommand{\EE}{\mathcal{E}}
\newcommand{\Cc}{\mathcal{C}}

\def\lm{\lambda_-}
\def\lp{\lambda_+}
\def\lpm{\lambda_\pm}

\def\Wt{\tau_\mathrm{W}}
\def\Ht{\tau_\mathrm{H}}
\def\Sm{\mathcal{S}}

\def\Ht{\tau_\mathrm{H}}
\def\Nc{N}
\def\Sm{\mathcal{S}}
\def\rho{\varrho}

\title[Top Eigenpair Statistics]{Top Eigenpair Statistics for Weighted Sparse Graphs}

\author{Vito A R Susca, Pierpaolo Vivo and Reimer K\"uhn}

\address{                 
King's College London, Department of Mathematics, Strand, London WC2R 2LS, United Kingdom
}

\begin{abstract}
We develop a formalism to compute the statistics of the top eigenpair of weighted sparse graphs with finite mean connectivity and bounded maximal degree. Framing the problem in terms of optimisation of a quadratic form on the sphere and introducing a fictitious temperature, we employ the cavity and replica methods to find the solution in terms of self-consistent equations for auxiliary probability density functions, which can be solved by population dynamics. This derivation allows us to identify and unpack the individual contributions to the top eigenvector's components coming from nodes of degree $k$. The analytical results are in perfect agreement with numerical diagonalisation of large (weighted) adjacency matrices, and are further cross-checked on the cases of random regular graphs and sparse Markov transition matrices for unbiased random walks.
\end{abstract}


\maketitle

\tableofcontents

\section{Introduction}

The largest eigenvalue and the associated \emph{top} eigenvector of a $N\times N$ matrix $J$ play a very important role in many applications. In multivariate data analysis and Principal Component Analysis, the top eigenpair of the covariance matrix provides information about the most relevant correlations hidden in the dataset \cite{Mardia1979, Monasson2015}. These extremal questions also arise in connection with synchronisation problems on networks \cite{Restrepo2005}, percolation problems \cite{Restrepo2008}, linear stability of coupled ODEs \cite{May1972}, financial stability \cite{Moran2019} and several other problems in physics and chemistry, connected to the applications of Perron's theorem \cite{MacCluer2000}. Also in the realm of quantum mechanics, the search for the ground state of a complicated Hamiltonian essentially amounts to solving the top eigenpair problem for a differential operator  \cite{Sakurai2017}. The top eigenpair is also relevant in signal reconstruction problems employing algorithms based on the spectral method \cite{Ma2019}.
In the context of graph theory, the eigenvectors of both adjacency and Laplacian matrices are employed to solve combinatorial optimisation problems, such as graph 3-colouring \cite{Alon1997} and to develop clustering and cutting techniques \cite{Coja-Oghlan2006,Pothen1990,Shi2000}. In particular, the top eigenvector of graphs is intimately related to the ``ranking" of the nodes of the network \cite{Brouwer2011}. Indeed, beyond the natural notion of ranking of a node given by its degree, the relevance of a node can be estimated from how ``important'' its neighbours are. The vector expressing the importance of each node is exactly the top eigenvector of the network adjacency matrix. Google PageRank algorithm works in a similar way \cite{Brin1998,Langville2008}: the PageRanks vector is indeed the top eigenvector of a large Markov transition matrix between web pages. 

When the matrix $J$ is random and symmetric with i.i.d. entries, analytical results on the statistics of the top eigenpair date back to the classical work by F\"uredi and Koml\'os \cite{Fueredi1981}: the largest eigenvalue of such matrices follows a Gaussian distribution with finite variance, provided that the moments of the distribution of the entries do not scale with the matrix size. This result directly relates to the largest eigenvalue of Erd\H{o}s-R\'enyi (E-R) \cite{Erdos1960} adjacency matrices in the case when the probability $p$ for two nodes to be connected does not scale with the matrix size $N$. This result has been then extended by Janson \cite{Janson2005} in the case when $p$ is large.  However, in our analysis we will be mostly dealing with the sparse case, i.e. when $p=c/N$, with $c$ being the constant mean degree of nodes (or equivalently, the mean number of nonzero elements per row of the corresponding adjacency matrix). In this sparse regime, Krivelevich and Sudakov \cite{Krivelevich2003} proved a theorem stating that for any constant $c$ the largest eigenvalue of Erd\H{o}s-R\'enyi graph diverges slowly with $N$ as $\sqrt{\log{N}/\log{ \log{N}}}$. To ensure that the largest eigenvalue remains $\sim\mathcal{O}(1)$, the nodes with very large degree must be pruned (see \cite{Ando2011}).

The characterisation of eigenvectors properties has proved to be much harder and is generally a less explored area of random matrix theory. Excluding the cases of i) invariant ensembles, where eigenvector components follow the celebrated Porter-Thomas distribution \cite{Porter1956,Livan2018}, ii) dense non-Hermitian matrices (see for instance the seminal works of Chalker and Mehlig \cite{Chalker1998} along with results about correlations between eigenvectors \cite{Janik1999,Fyodorov2018} and some more recent applications \cite{Burda2015,Nowak2018,Gudowska-Nowak2018,NeriMetz2019}) and iii) perturbed matrices \cite{Truong2016,Facoetti2016,Burda2017,Allez2012,Bun2018}, systematic results are scarcer for sparse Hermitian matrices, especially in the limit of high sparsity. Indeed, although Gaussian statistics and delocalisation of eigenvectors are known properties of adjacency matrices of Erd\H{o}s-R\'enyi and random regular graphs in the case where the mean degree $c=c(N)$ diverges with $N$ \cite{Tran2013, Bourgade2017, Dumitriu2012}, very few results are available for the high sparsity regime, i.e. with fixed $c$. In this limit, numerical studies have shown that most of the eigenvectors of a random regular graph follow a Gaussian distribution \cite{Elon2008}, as well as almost-eigenvectors \cite{Backhausz2016}, whereas Erd\H{o}s-R\'enyi eigenvectors are localised especially for low values of $c$. 

The statistics of the first eigenvector components for very sparse symmetric random matrices was first considered in the seminal works by Kabashima and collaborators \cite{Kabashima2010, Kabashima2012, Takahashi2014}, which constitute the starting point of our analysis. The focus there is on specific classes of real sparse random matrices, i.e. when the matrix connectivity is either a random regular graph or a mixture of multiple degrees, and the nonzero elements are drawn from a Bernoulli
distribution. More precisely, in \cite{Kabashima2010} and \cite{Takahashi2014} the cavity method was employed for the top eigenpair problem, while in \cite{Kabashima2012} the replica formalism was instead adopted to study the same problem in the thermodynamic limit, recovering cavity results. Our aim is to analyse and develop both the cavity and replica formalisms they pioneered even further, and to present them in a unified way that looks - at least to our eyes - more transparent.

We will be implementing a Statistical Mechanics formulation of the top eigenpair problem, using both the cavity (Section \ref{sec:cavity}) and replica (Section \ref{sec:replica}) methods - borrowed from the standard arsenal of disordered systems physics - as main solving tools. 

The replica method, widely used in the physics of spin glasses \cite{Zamponi2010}, was first introduced in the context of random matrices by Edwards and Jones \cite{Edwards1976} to compute the average spectral density of random matrices defined in terms of the joint probability density function (pdf) of their entries. Building on this approach, Bray and Rodgers in their seminal paper \cite{Rodgers1988} were able to express the spectral density of Erd\H{o}s-R\'enyi adjacency matrices as the solution of a (nearly intractable) integral equation. Therefore, asymptotic analyses for large average connectivities \cite{Rodgers1988}, and approximation schemes such as the single defect approximation (SDA) and the effective medium approximation (EMA) \cite{Biroli1999, Semerjian2002} were first developed as a way around this hindrance. 
An alternative approach was pursued in \cite{Kuehn2008} (see also \cite{Bianconi2008}): starting from Bray-Rodgers replica-symmetric setup \cite{Rodgers1988}, the functional order parameters of the theory are expressed as continuous superpositions of Gaussians with fluctuating variances, as suggested by earlier solutions of models for finitely coordinated harmonically coupled systems \cite{Kuehn2007}. This formulation gives rise to non-linear integral equations for the probability densities of such variances, which can be efficiently solved by a population dynamics algorithm. Our paper will follow a similar approach in Section \ref{sec:replica}.
 
The cavity method \cite{Mezard1987}, also known as Bethe-Peierls or belief-propagation method, was introduced in the context of disordered systems and sparse random matrices as a more intuitive and straightforward alternative to replicas: the two methods are known to provide the same results for the spectral density of graphs \cite{Slanina2011}, even though a general, first-principle proof of their equivalence does not seem to be currently available. A rigorous proof of the correctness of cavity method and the tree-like approximation for finitely coordinated graphs is given in \cite{Bordenave2010}. One of the advantages of the cavity method is that it allows one to solve the spectral problem for very large single instances of sparse random graphs, as done in \cite{Rogers2008}.
Both the replica and cavity approaches in \cite{Kuehn2008} and \cite{Rogers2008} retrieve known results such as the Kesten-McKay law for the spectra of random regular graphs \cite{Kesten1959, McKay1981}, the Mar\v{c}enko-Pastur law and the Wigner's semicircle law respectively for sparse covariance matrices and for Erd\H{o}s-R\'enyi adjacency matrices in the large mean degree limit. Both approaches have also been employed to characterise the spectral density of sparse Markov matrices \cite{Kuehn2015, Kuehn2015a} and graphs with modular \cite{Erguen2009} and small-world \cite{Kuehn2011} structure and with topological constraints \cite{Rogers2010}. The localisation transition for sparse symmetric matrices was studied in \cite{Metz2010}. The two methods have also been extended to the study of the spectral density of sparse non-Hermitian matrices \cite{Rogers2009,Neri2012}, whereas eigenvalue outliers have been considered in \cite{Neri2016}; for an excellent review, see \cite{Metz2018}. The spectral properties of the Hashimoto non-backtracking operator - arising in the cavity solution (see \ref{sec:Non_backtracking} for details) have  been investigated in \cite{Saade2014,Bordenave2015,Bordenave2016}. In this paper, we propose a ``grand canonical" cavity derivation that differs in the details from \cite{Kabashima2010} (see Section \ref{sec:cavity}). We also provide a detailed analysis of the single-instance recursion equations, showing that their convergence is strictly related to the spectral properties of a modified non-backtracking operator associated with the single-instance matrix. At the same time, building on the insights coming from the replica treatment, we are able to better understand the behaviour of the stochastic recursions that provide the solution of the top eigenpair problem in the thermodynamic limit. Furthermore, the population dynamics algorithm employed to solve these recursions allows us to characterise the distributions of the cavity fields in the thermodynamic limit and identify the individual contributions of nodes of different degrees $k$ to the top eigenvector's entries. 

The plan of the paper is as follows. In Section \ref{Formulation}, we will formulate the problem and provide the main starting points. In Section \ref{sec:cavity}, we will describe the cavity approach to the problem, first for the single instance case (in \ref{cavity_singleinstance}), and then in the thermodynamic limit (in \ref{cavity_therm}). In Section \ref{sec:replica}, we formulate the replica approach to the same problem, first focussing on the largest eigenvalue problem (in \ref{sec:replica_eigenvalue}) and then on the density of top eigenvector's components (in \ref{sec:replica_eigenvector}). For both problems, we take the weighted Erd\H{o}s-R\'enyi and random regular graphs as representative examples. In Section \ref{sec:Markov} we build on our previous results to complete the picture for Markov transition matrices on a random graph structure. In Section \ref{sec:Population-Dynamics}, we provide the details of the population dynamics algorithm, and in Section \ref{sec:conclusions} we offer a summary and outlook for future research. In \ref{sec:Non_backtracking}, we provide a detailed discussion of the single-instance cavity approach and associated non-backtracking operator. In \ref{sec:No_shortcuts}, we offer a detailed replica derivation of the typical location of the largest eigenvalue for sparse graphs characterised by a generic degree distribution $p(k)$.

\section{Formulation of the problem} \label{Formulation}

We consider a sparse random $N\times N$ symmetric matrix $J=\left(J_{ij}\right)$,
with real i.i.d. entries. The matrix entries are defined as

\begin{equation}
J_{ij}=c_{ij}K_{ij}\,,\label{eq:matrix_def}
\end{equation}
where the $c_{ij}\in\{0,1\}$ constitute the connectivity matrix, i.e. the adjacency
matrix of the underlying graph, and the $K_{ij}$ encode bond weights. We will typically consider the case of Poissonian highly sparse connectivity - where the node degrees $k_i$ (or equivalently the number of nonzero elements per row of $J$) fluctuate according to a bounded Poisson distribution
\begin{equation}
P(k_i=k)=\mathcal{N}^{-1}\mathrm{e}^{-\bar{c}}\bar{c}^k/k!\ ,\qquad k=0,\ldots,k_{\mathrm{max}}\ ,\label{eq:degree_pmf}
\end{equation}
 with the mean degree a finite constant $c\equiv\left\langle k\right\rangle$ and $\mathcal{N}=\sum_{k=0}^{k_\mathrm{max}}\mathrm{e}^{-\bar{c}}\bar{c}^k/k!$ to ensure normalisation. The bond weights $K_{ij}$ will be i.i.d. random variables drawn from a parent pdf $p(K)$ with bounded support. This setting is sufficient to ensure that the largest eigenvalue $\lambda_1$ of $J$ will remain of $\mathcal{O}(1)$ for $N\to\infty$.

The spectral theorem ensures that $J$ can be diagonalised via an orthonormal basis of
eigenvectors $\bm{v}_{\alpha}$ with corresponding real eigenvalues
$\lambda_{\alpha}$ ($\alpha=1,\ldots,N$),

\begin{equation}
J\bm{v}_{\alpha}=\lambda_{\alpha}\bm{v}_{\alpha}\ ,
\end{equation}
for each eigenpair $\alpha=1,\ldots,N$. We assume that there is no eigenvalue degeneracy, and that they are sorted $\lambda_1>\lambda_2>\ldots >\lambda_N$.

The goal of this work is to set up a formalism based on the statistical mechanics of disordered systems to find:
\begin{itemize}
\item The average (or typical value) $\langle \lambda_1\rangle_J$ of the largest eigenvalue $\lambda_1$.
\item The density $\rho(u)=\Big\langle\frac{1}{N}\sum_{i=1}^N\delta(u-v_1^{(i)}) \Big\rangle_J$ of the top eigenvector's components, $\bm v_1 =(v_1^{(1)},\ldots,v_1^{(N)})$ ,
\end{itemize}
where the average $\langle\cdot\rangle_J$ is taken over the distribution of the matrix $J$.

The problem can be formulated as the optimisation problem
of a quadratic function $\hat{H}(\bm{v})$, according to which $\bm{v}_{1}$ is
the vector normalized to $N$ that realises the condition

\begin{equation}
N\lambda_1=\min_{| \bm{v}|^2 =N}\left[ \hat{H}(\bm{v})\right] =\min_{| \bm{v}|^2 =N}\left[-\frac{1}{2}\left(\bm{v},J\bm{v}\right)\right]\ ,\label{eq:minimum}
\end{equation}
as dictated by the Courant-Fischer definition of eigenvectors. The round
brackets $\left(\cdot,\cdot\right)$ indicate the dot product between
vectors in $\mathbb{R}^{N}$. It is easy to show that $\hat{H}\left(\bm{v}\right)$ is bounded
\begin{equation}
-\frac{1}{2}\lambda_{1}N\leq\hat{H}\left(\bm{v}\right)\leq-\frac{1}{2}\lambda_{N}N\,,
\end{equation}
and attains its minimum
when computed on the top eigenvector.

For a fixed matrix $J$, the minimum in \eqref{eq:minimum} can be computed by introducing a fictitious canonical ensemble of $N$-dimensional vectors $\bm v$ at inverse temperature $\beta$, whose Gibbs-Boltzmann distribution reads
\begin{equation}
P_{\beta,J}(\bm v)=\frac{1}{Z}\exp\left[\frac{\beta}{2}(\bm v, J\bm v)\right]\delta(|\bm v|^2-N)\ ,\label{eq:hard}
\end{equation}
where the delta function enforces normalisation. Clearly, in the low temperature limit $\beta\to\infty$, only one 'state' remains populated, which corresponds to $\bm v=\bm v_1$, the top eigenvector of the matrix $J$. The hard normalisation constraint can also be relaxed for our purposes, and replaced with a soft, "grand canonical'' version
\begin{equation}
P_{\beta,J}\left(\bm{v}\right)=\frac{1}{Z}\exp\left\{ \beta\left[\frac{1}{2}\left(\bm{v},J\bm{v}\right)-\frac{\lambda}{2}\left(\bm{v},\bm{v}\right)\right]\right\} \ ,\label{eq:boltz_pdf}
\end{equation}
where $\lambda $ is an auxiliary Lagrange multiplier. The two versions above are expected to provide the same physical results in the limits $\beta,N\to\infty$, as we explicitly demonstrate by using \eqref{eq:boltz_pdf} for our \emph{cavity}  treatment  in Section \ref{sec:cavity}, and \eqref{eq:hard} as a starting point of our \emph{replica} calculation in Section \ref{sec:replica}.

\section{Cavity approach \label{sec:cavity}}
In what follows, we will use a cavity method formulation for the top eigenpair problem which is deeply rooted in the statistical mechanics approach to disordered systems. Our formulation provides equations for the statistics of the top eigenpair that are fully equivalent to those found earlier by Kabashima et al. in \cite{Kabashima2010}. Our treatment, however, brings more neatly to the surface a few subtleties related to the solution of self-consistency equations and their range of applicability, this way providing a more transparent derivation. 

The central idea of the cavity method \cite{Mezard1987} consists in computing observables related to a given node, relying on some information concerning its neighbourhood \emph{when the node of interest is removed from the network}. It is useful every time the underlying graph has a finite connectivity structure: its predictions become exact for trees and approximately exact for tree-like structures (where loops are negligible) such as graphs in the high sparsity regime.

\begin{figure}
\begin{centering}
\includegraphics[scale=1]{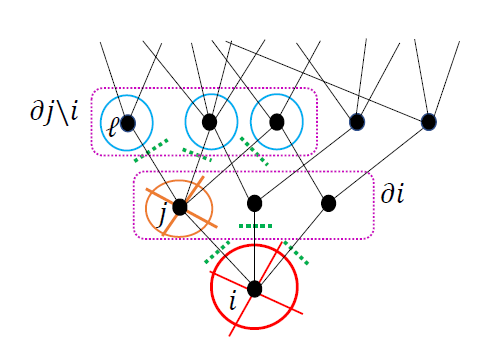}
\par\end{centering}
\begin{centering}
\caption{\label{cavity_tree}Tree-like structure of a graph. The indexing refers to the labels used in the cavity method treatment in subsection \ref{cavity_singleinstance}.}
\par\end{centering}
\centering{}
\end{figure}

\subsection{Single instance\label{cavity_singleinstance}}

Consider for the time being a single instance of the random matrix $J$. Starting from the soft-constraint distribution \eqref{eq:boltz_pdf}, whose partition function is
\begin{equation}
Z=\int\mathrm{d}\bm{v}\exp\left\{ \beta\left[\frac{1}{2}\left(\bm{v},J\bm{v}\right)-\frac{\lambda}{2}\left(\bm{v},\bm{v}\right)\right]\right\} \ ,\label{eq:cavity_partition}
\end{equation}
it is trivial to notice that the condition $\lambda>\lambda_1$ is necessary to ensure convergence for all $\beta$.

The marginal distribution of the component $v_{i}$, obtained by integrating out all other components in \eqref{eq:boltz_pdf}, and using the sparsity condition $J_{ij}=0$ if $j\notin\partial i$ (where $\partial i$ denotes the immediate neighbourhood of $i$) is

\begin{equation}
P_{i}\left(v_{i}\right)=\frac{1}{Z_{i}}\exp\left( -\beta\frac{\lambda}{2}v_{i}^{2}\right) \int\mathrm{d}\bm{v}_{\partial i}\exp\left( \beta\sum_{j\in\partial i}J_{ij}v_{i}v_{j}\right) P^{(i)}\left(\bm{v}_{\partial i}\right)\ ,\label{eq:marginal}
\end{equation}
where $P^{(i)}(\bm{v}_{\partial i})$ is the joint distribution of the components pertaining to the immediate neighbourhood of $i$, $\partial i$, when the node $i$ has been removed. Indeed, all the components outside $\partial i$ can be integrated out without difficulty, and the resulting constant term can be just reabsorbed in the normalisation constant. $P^{(i)}(\bm{v}_{\partial i})$ is also known as \textit{cavity probability
distribution}. 

Adopting now a tree-like approximation, which is accurate for very sparse graphs, all nodes $j$ in $\partial i$ are connected with each other only through $i$ (see Fig. \ref{cavity_tree}), therefore they get disconnected when the node $i$ is removed from the network: this implies that the integral appearing in \eqref{eq:marginal}
factorises as

\begin{equation}
P_{i}\left(v_{i}\right)=\frac{1}{Z_{i}}\exp\left( -\beta\frac{\lambda}{2}v_{i}^{2}\right) \prod_{j\in\partial i}\int\mathrm{d}v_{j}\exp\left(\beta J_{ij}v_{i}v_{j}\right) P_{j}^{(i)}\left(v_{j}\right)\,.\label{eq:marginal_2}
\end{equation}

In the same way, a similar expression can be derived for the marginal
cavity distribution $P_{j}^{(i)}(v_{j})$ now appearing in \eqref{eq:marginal_2}.
Iterating the reasoning as before, and further removing the node $j\in\partial i$
in the network in which the node $i$ had already been removed, one
can write

\begin{equation}
P_{j}^{(i)}\left(v_{j}\right)=\frac{1}{Z_{j}^{(i)}}\exp\left( -\beta\frac{\lambda}{2}v_{j}^{2}\right) \prod_{\ell\in\partial j\backslash i}\int\mathrm{d}v_{\ell}\exp\left( \beta J_{j\ell}v_{j}v_{\ell}\right) P_{\ell}^{(j)}\left(v_{\ell}\right)\ ,\label{eq:cavity_marg}
\end{equation}
where the symbol $\partial j\backslash i$ denotes the neighbourhood of $j$ excluding $i$.

Equation \eqref{eq:cavity_marg} has now become a self-consistent equation for the cavity probability distributions,
which can be solved by a Gaussian ansatz for $P_{j}^{(i)}\left(v_{j}\right)$ 

\begin{equation}
P_{j}^{(i)}\left(v_{j}\right)=\sqrt{\frac{\beta \Omega_{j}^{(i)}}{2\pi}}\exp\left(- \frac{\beta {H_{j}^{(i)}}^2}{2\Omega_{j}^{(i)}}\right)\exp\left( -\frac{\beta}{2}\Omega_{j}^{(i)}v_{j}^{2}+\beta H_{j}^{(i)}v_{j}\right)\ , \label{eq:gaussian_ansatz}
\end{equation}
where the parameters $\Omega_{i}^{(i)}$ and $H_{j}^{(i)}$ are called
\textit{cavity fields}. This ansatz is chosen to obtain
a solution $\bm{v}$ whose components are not peaked at zero in
the $\beta\rightarrow\infty$ limit. Inserting the Gaussian ansatz
\eqref{eq:gaussian_ansatz} in \eqref{eq:cavity_marg} and performing
the resulting Gaussian integrals, one obtains

\begin{equation}
P_{j}^{(i)}\left(v_{j}\right)=\frac{1}{Z_{j}^{(i)}}\exp\left( -\frac{\beta}{2}\lambda v_{j}^{2}\right) \prod_{\ell\in\partial j \backslash i}\exp\left[ \frac{\beta}{2}\frac{\left(J_{j\ell}v_{j}+H_{\ell}^{(j)}\right)^{2}}{\Omega_{\ell}^{(j)}}\right] \ .\label{eq:gaussian_ansatz2}
\end{equation}
Comparing the coefficients of the same powers of $v_j$ between \eqref{eq:gaussian_ansatz} and \eqref{eq:gaussian_ansatz2}, we obtain the following two self-consistent relations which
define the cavity fields $\Omega_{i}^{(i)}$ and $H_{j}^{(i)}$ 

\begin{align}
\Omega_{j}^{(i)} &=\lambda-\sum_{\ell\in\partial j \backslash i}\frac{J_{j\ell}^{2}}{\Omega_{\ell}^{(j)}}\ ,\label{eq:A}\\ 
H_{j}^{(i)} &=\sum_{\ell\in\partial j \backslash i}\frac{J_{j\ell}}{\Omega_{\ell}^{(j)}}H_{\ell}^{(j)}\ .\label{eq:H}
\end{align}
These equations have been obtained before in \cite{Kabashima2010}. 

The Gaussian ansatz \eqref{eq:gaussian_ansatz} can then be inserted
in \eqref{eq:marginal_2}, resulting in a Gaussian distribution
for the single-site marginals

\begin{equation}
P_{i}\left(v_{i}\right)=\frac{1}{Z_i}\exp\left( -\frac{\beta}{2}\Omega_{i}v_{i}^{2}+\beta H_{i}v_{i}\right) \,,\label{eq:single_site_marginals}
\end{equation}
where the $N$ coefficients $\Omega_{i}$ and $H_{i}$ are given by the following equations

\begin{align}
\Omega_{i} &=\lambda-\sum_{j\in\partial i}\frac{J_{ij}^{2}}{\Omega_{j}^{(i)}}\ ,\label{eq:A_marg}\\
H_{i} &=\sum_{j\in\partial i}\frac{J_{ij}}{\Omega_{j}^{(i)}}H_{j}^{(i)}\ .\label{eq:H_marg}
\end{align}
Here, $\Omega_j^{(i)}$ and $H_j^{(i)}$ are the fixed-point solutions of \eqref{eq:A} and \eqref{eq:H}.

In the limit $\beta\to\infty$, the marginal distribution \eqref{eq:single_site_marginals} converges to
\begin{equation}
P_i(v_i)=\delta\left(v_{i}-\frac{H_{i}}{\Omega_{i}}\right)\ ,\label{eq:components_single}
\end{equation}
from which one concludes that the components of the top eigenvector of the fixed matrix $J$ (a single instance of the ensemble) must be given by $v_1^{(i)}=H_{i}/\Omega_{i}$, where $H_{i}$ and $\Omega_{i}$ are the values obtained from \eqref{eq:A_marg} and \eqref{eq:H_marg}, after the fixed-points of the recursions \eqref{eq:A} and \eqref{eq:H} have been obtained.

A detailed discussion on how to solve the above recursions in practice and on the role of the (yet unspecified) multiplier $\lambda$ is deferred to  \ref{sec:Non_backtracking}. Although this derivation only relies on the tree-like approximation for the local connectivity and is arguably very easy and intuitive, it is not particularly interesting as it stands: the complexity of the cavity algorithm for a single instance is actually \emph{higher} than a high-precision, direct diagonalisation of the matrix $J$, therefore it is of little practical use \emph{per se}. It is, however, a conceptually necessary ingredient to discuss infinite-size matrices, as we do in the next subsection.

\subsection{Thermodynamic limit $N\to\infty$}\label{cavity_therm}

In an infinitely large network, it is no longer possible to keep track of an infinite number of cavity fields. Following \cite{Kabashima2010}, we consider first the \emph{joint probability density} that the cavity fields of type $\Omega_{j}^{(i)}$ and $H_{j}^{(i)}$ take up values around $\omega$ and $h$

\begin{align}
\nonumber q\left(\omega,h\right)&=\mathrm{Prob}\left(\Omega_{j}^{(i)}=\omega,H_{j}^{(i)}=h\right) \\
&=\left(\sum_{i=1}^{N}k_{i}\right)^{-1}\sum_{i=1}^{N}\sum_{j\in\partial i}\delta\left(\omega-\Omega_{j}^{(i)}\right)\delta\left(h-H_{j}^{(i)}\right)\ ,
\end{align}
where $N$ is now large but finite. This is a properly normalised pdf: indeed, we can associate two cavity fields $\Omega_{j}^{(i)}$ and $H_{j}^{(i)}$ to any link $(i,j)$ of the network. Since every node $i$ is the source of $k_i$ links, their total number is given by $\sum_{i=1}^{N}k_{i}$.

Next, one may appeal to the single-instance update rules given by \eqref{eq:A} and \eqref{eq:H} to characterise the above distribution self-consistently, as is done in \cite{Kabashima2010}.
It should be stressed that in an infinitely large network links can only be distinguished by the degree of the node they are pointing
to. Thus, for a given edge $(i,j)$ pointing to a node $j$ of degree $k$, the values $\omega$ and $h$ of the pair of cavity fields $\Omega_{j}^{(i)}$ and $H_{j}^{(i)}$ living on this edge are determined respectively by the $k-1$ values $\left\{\omega_{\ell}\right\}$ and $\left\{h_{\ell}\right\}$ of the cavity fields $\Omega_{\ell}^{(j)}$ and $H_{\ell}^{(j)}$ living on each of the edges connecting $j$ with its neighbours $\ell\in\partial j \backslash i$. In an infinite system, these values can be thought of as $k-1$ independent realisations of the random variables of types $\Omega_{j}^{(i)}$ and $H_{j}^{(i)}$, drawn from their joint pdf $q(\omega,h)$. The entries of $J$ that appear in the single instance recursions \eqref{eq:A} and \eqref{eq:H} are replaced by a set $\left\{K_\ell\right\}_{k-1}$ of $k-1$  independent realisations of the random variable $K$, each distributed according to the pdf  $p(K)$ of bond weights. The full distribution $q(\omega,h)$ is then obtained by weighing each edge contribution with the probability $r(k)$ of having a random link pointing to a node of degree $k$ and summing up over all possible degrees up to $k_{\mathrm{max}}$, leading to the self-consistency equation

\begin{equation}
q\left(\omega,h\right)=\sum_{k=1}^{k_{\mathrm{max}}}r\left(k\right)\!\!\int\left[\prod_{\ell=1}^{k-1}\mathrm{d}q\left(\omega_{\ell},h_{\ell}\right)\right]\left\langle \delta\left(\omega-\lambda+\sum_{\ell=1}^{k-1}\frac{K_{\ell}^{2}}{\omega_{\ell}}\right)\delta\left(h-\sum_{\ell=1}^{k-1}\frac{h_{\ell}K_{\ell}}{\omega_{\ell}}\right)\right\rangle _{{\{K\}}_{k-1}}\ ,\label{eq:q}
\end{equation}
where $\mathrm{d}q\left(\omega_{\ell},h_{\ell}\right)\equiv\mathrm{d}\omega_\ell\mathrm{d}h_\ell q\left(\omega_{\ell},h_{\ell}\right)$, and the average $\langle\cdot\rangle_{{\{K\}}_{k-1}}$ is taken over $k-1$ independent realisations of the random variable $K$. We recall that 

\begin{equation}
r\left(k\right)=\frac{kp\left(k\right)}{\left\langle k \right\rangle}\,,
\label{eq:pmf_link}
\end{equation}
 where $p\left(k\right)$ is the probability of having a node of degree
$k$ and $\left\langle k \right\rangle=\sum_{k}kp\left(k\right)$ \cite{Dorogovtsev2002}. The sum in \eqref{eq:q}  starts from $k=1$ since we should not be concerned with isolated nodes.

Eq. \eqref{eq:q} is generally solved via a population dynamics algorithm
(see Section \ref{sec:Population-Dynamics} for details). In some exceptional cases, such as for adjacency matrices of random
regular graphs, it can be solved analytically (see discussion in sections \ref{sec:rrg} and \ref{sec:rrg2} below).

In a similar fashion, the joint pdf of the
coefficients $\Omega_{i}$ and $H_{i}$ can be expressed as

\begin{equation}
Q\left(\Omega,H\right)=\frac{1}{N}\sum_{i=1}^{N}\delta\left(\Omega-\Omega_{i}\right)\delta\left(H-H_{i}\right)\ .
\end{equation}
In this case, there is a pair of marginal coefficients $\Omega_{i}$ and $H_{i}$ living on each \emph{node}. Since in the infinite size limit the nodes can only be distinguished by their degree, following the same line of reasoning that led to \eqref{eq:q}, the joint pdf of the random variables of the type $\Omega_{i}$ and $H_{i}$  in the thermodynamic limit can be written as
 
\begin{equation}
Q\left(\Omega,H\right)=\sum_{k=0}^{k_{\mathrm{max}}}p\left(k\right)\!\!\int\left[\prod_{\ell=1}^{k}\mathrm{d}q\left(\omega_{\ell},h_{\ell}\right)\right]\left\langle \delta\left(\Omega-\lambda+\sum_{\ell=1}^{k}\frac{K_{\ell}^{2}}{\omega_{\ell}}\right)\delta\left(H-\sum_{\ell=1}^{k}\frac{h_{\ell}K_{\ell}}{\omega_{\ell}}\right)\right\rangle _{{\{K\}}_k}\ ,\label{eq:Q}
\end{equation}
where $p(k)$ is the degree distribution. Here, $q\left(\omega_{\ell},h_{\ell}\right)$ is the fixed-point distribution of cavity fields, i.e. the solution of the self-consistency equation \eqref{eq:q}, which should therefore be solved beforehand.

The distribution of the top eigenvector's components in the thermodynamic limit is then
obtained in terms of the pdf $Q\left(\Omega,H\right)$ in \eqref{eq:Q}, exploiting the analogy with the single-instance case in \eqref{eq:components_single}, and reads

\begin{equation}
\rho(u)=\left\langle \frac{1}{N}\sum_{i=1}^{N}\delta\left(u-v_{1}^{(i)}\right)\right\rangle =\int\mathrm{d}\Omega\mathrm{d}H~Q\left(\Omega,H\right)\delta\left(u-\frac{H}{\Omega}\right)\ .\label{eq:first_eigv_comp_denisity_cavity}
\end{equation}

Both equations \eqref{eq:q} and \eqref{eq:Q} still depend on the parameter $\lambda$: it must be fixed taking into account the normalisation of the top eigenvector. This condition amounts to requiring that
\begin{equation}
1=\langle u^2\rangle=\int\mathrm{d}\Omega\mathrm{d}H~Q\left(\Omega,H\right)\frac{H^2}{\Omega^2}\ .\label{eq:cavity_normalization}
\end{equation}
Crucially, the value of $\lambda$ for which the above normalisation condition is satisfied turns out to be exactly equal to the typical largest eigenvalue, $\lambda\equiv\left\langle \lambda_{1}\right\rangle_J$. Indeed, for every $\lambda>\left\langle \lambda_{1}\right\rangle_J$, the distribution of the $h$'s shrinks to a delta peak located at zero, whereas for $\lambda<\left\langle \lambda_{1}\right\rangle_J$, negative values of the $\omega$'s start to appear while the $h$'s grow without bounds in the self-consistency solution of \eqref{eq:q}. This is not surprising, since $\lambda<\left\langle \lambda_{1}\right\rangle_J$ is precisely the range of values for $\lambda$ that makes the Gibbs-Boltzmann distribution \eqref{eq:boltz_pdf} not normalisable. 

As a final remark on the cavity solution, the equations \eqref{eq:q} and \eqref{eq:first_eigv_comp_denisity_cavity} will match respectively \eqref{eq:pi_trunc} and \eqref{eq:density_with_pi} obtained via the replica method in Section \ref{sec:replica} below.

The discussion above has the advantage of leading rather quickly to the results \eqref{eq:Q} and \eqref{eq:first_eigv_comp_denisity_cavity}. It is, however, instructive to reconsider this problem from the point of view of the replica approach, which provides a lengthier but rather systematic procedure, and arrives at the very same equations while departing from very different premises. Both approaches (cavity or replicas) present different advantages and drawbacks - especially if seen through the prism of full mathematical rigour - and it is therefore of interest to compare them back to back. For the sake of clarity, we will keep the two pathways (typical largest eigenvalue vs. density of top eigenvector's components) clearly separate until the point where we realise that the same self-consistency equation governs the statistics of both quantities.

\section{\label{sec:replica}Replica derivation}

In this section, we evaluate  the average location of the largest eigenvalue and the density of top eigenvectors' components within the replica framework. The starting point of our analysis is the formalism pioneered in \cite{Kabashima2012}. However, our derivation is not confined to specific connectivity distributions of the matrix entries as in \cite{Kabashima2012}, and thus provides a rather general and robust methodology that can be applied to any graph with finite mean connectivity and bounded maximal degree. We also make a quite transparent and convincing case for the equivalence between the cavity and replica methods in these problems. Moreover, as we did for the cavity approach, we thoroughly discuss bounds on the values of parameters that guarantee a converging solution.

\subsection{Typical largest eigenvalue}\label{sec:replica_eigenvalue}

Consider again a $N\times N$ symmetric matrix $J_{ij}=c_{ij}K_{ij}$. The joint distribution of the matrix entries is

\begin{equation}
P\left(\left\{ J_{ij}\right\} \middle | \left\{k_i\right\} \right)=P\left(\left\{c_{ij}\right\} \middle | \left\{k_i\right\} \right)\prod_{i<j}\delta_{K_{ij},K_{ji}}p\left(K_{ij}\right)\ ,\label{eq:joint_true}
\end{equation}
where, in the framework of the configuration model \cite{Kuehn2011}, the distribution $P\left(\left\{c_{ij}\right\} \middle | \left\{k_i\right\} \right)$ of connectivities $\left\{c_{ij}\right\}$ compatible with a given degree sequence $\left\{k_i\right\}$ is given by
\begin{equation}
P\left(\left\{ c_{ij}\right\} \middle | \left\{k_i\right\} \right)=\frac{1}{\mathcal{M}}\prod_{i<j}\delta_{c_{ij},c_{ji}}\left(\frac{c}{N}\delta_{c_{ij},1}+\left(1-\frac{c}{N}\right)\delta_{c_{ij},0}\right)\prod_{i=1}^{N}\delta_{\sum_{j}c_{ij},k_{i}}\ ,\label{eq:boundedER_connectivity}
\end{equation}
and the pdf $p\left(K_{ij}\right)$ of bond weights (with compact support and upper edge $\zeta$) can be kept unspecified until the very end. 

It has been shown in many works \cite{Kuehn2008,Kuehn2015a} that a convenient shortcut for the calculation consists in replacing the ``microcanonical" Eq. \eqref{eq:boundedER_connectivity} with the standard Erd\H{o}s-R\'enyi connectivity distribution

\begin{equation}
P\left(\left\{ c_{ij}\right\} \right)=\prod_{i<j}\delta_{c_{ij},c_{ji}}\left(\frac{c}{N}\delta_{c_{ij},1}+\left(1-\frac{c}{N}\right)\delta_{c_{ij},0}\right)\ .\label{eq:ER_connectivity}
\end{equation}
Although Eq. \eqref{eq:ER_connectivity} technically gives rise to an \emph{unbounded} Poisson degree distribution with mean $c$ -- and therefore a largest eigenvalue whose location typically \emph{grows} with $N$ \cite{Krivelevich2003} --  the final results (e.g. Eq. \eqref{eq:pi}) can be easily adjusted and extended to cover \emph{any} degree distribution $p(k)$ with finite mean and bounded largest degree. 
For simplicity, we will therefore consider the distribution of the matrix entries to be simply

\begin{equation}
P\left( \left\{ J_{ij}\right\} \right)=P\left(\left\{c_{ij}\right\} \right)\prod_{i<j}\delta_{K_{ij},K_{ji}}p\left(K_{ij}\right)\ 
\label{eq:joint_er}
\end{equation} 
at the outset, where $P\left( \left\{ c_{ij}\right\} \right)$ is given by \eqref{eq:ER_connectivity}. Once the Erd\H{o}s-R\'enyi  Poissonian degree distribution has appeared in the formulae, it will be straightforward to replace it with the actual finite-mean degree distribution of interest (for instance, the \emph{truncated} Poisson distribution \eqref{eq:degree_pmf}). In \ref{sec:No_shortcuts}, we will however provide a first-principle derivation for sparse graphs with a generic degree distribution $p(k)$, without relying on any shortcut.

The average of the largest eigenvalue can be computed as the formal limit
\begin{equation}
\left\langle \lambda_{1}\right\rangle_J=\lim_{\beta\rightarrow\infty}\frac{2}{\beta N}\left\langle \ln Z\right\rangle _{J},\qquad Z=\int\mathrm{d}\bm{v}\exp\left[ \frac{\beta}{2}\left(\bm{v},J\bm{v}\right)\right] \delta\left(\left|\bm{v}\right|^{2}-N\right)\ ,\label{eq:average_lambda1summary}
\end{equation}
in terms of the quenched free energy of the model defined in \eqref{eq:hard}.

The average over $J$ is computed using the replica
trick as follows

\begin{equation}
\left\langle \lambda_{1}\right\rangle _{J}=\lim_{\beta\rightarrow\infty}\frac{2}{\beta N}\lim_{n\rightarrow0}\frac{1}{n}\ln\left\langle Z^{n}\right\rangle _{J}\ ,\label{formulalargest}
\end{equation}
where $n$ is initially taken as an integer, and then analytically continued to real values in the vicinity of $n=0$.

The replicated partition function is

\begin{equation}
\left\langle Z^{n}\right\rangle _{J}=\int\left(\prod_{a=1}^{n}\mathrm{d}\bm{v}_{a}\right)\left\langle \exp\left( \frac{\beta}{2}\sum_{a=1}^{n}\sum_{i,j}^{N}v_{ia}J_{ij}v_{ja}\right) \right\rangle _{J}\ \prod_{a=1}^{n}\delta\left(\left|\bm{v}_{a}\right|^{2}-N\right)\,. \label{eq:rep_Z}
\end{equation}

Taking the average w.r.t the joint distribution \eqref{eq:joint_er} of matrix entries yields \cite{Kuehn2008}

\begin{equation}
\left\langle \exp\left( \frac{\beta}{2}\sum_{a=1}^{n}\sum_{i,j}^{N}v_{ia}J_{ij}v_{ja}\right) \right\rangle _{J}=\exp\left[ \frac{c}{2N}\sum_{i,j}\left(\left\langle \mathrm{e}^{\beta K\sum_{a}v_{ia}v_{ja}}\right\rangle _{K}-1\right)\right]\ ,
\end{equation}
where $\left\langle\cdot \right\rangle _{K}$ denotes averaging over
the single-variable pdf $p\left(K\right)$ characterising
the i.i.d. bond weights $K_{ij}$.

We also employ a Fourier representation of the Dirac delta enforcing
the normalisation constraints

\begin{equation}
\prod_{a=1}^{n}\delta\left(\left|\bm{v}_{a}\right|^{2}-N\right)=\int_{-\infty}^{\infty}\left(\prod_{a=1}^{n}\frac{\beta}{2}\frac{\mathrm{d}\lambda_{a}}{2\pi}\right)\prod_{a=1}^{n}\exp\left[ -\mathrm{i}\frac{\beta}{2}\lambda_{a}\left(\sum_{i=1}^{N}v_{ia}^{2}-N\right)\right] \,.
\end{equation}

The replicated partition function thus becomes

\begin{align}
\nonumber\left\langle Z^{n}\right\rangle _{J} & =\left(\frac{\beta}{4\pi}\right)^{n}\int\left(\prod_{a=1}^{n}\mathrm{d}\bm{v}_{a}\mathrm{d}\lambda_a\right)\exp\left( \mathrm{i}\frac{\beta}{2}N\sum_{a}\lambda_{a}\right) \exp\left( -\mathrm{i}\frac{\beta}{2}\sum_{a}\sum_{i}\lambda_{a}v_{ia}^{2}\right)\\
 & \times\exp\left[ \frac{c}{2N}\sum_{i,j}\left(\left\langle \mathrm{e}^{\beta K\sum_{a}v_{ia}v_{ja}}\right\rangle _{K}-1\right)\right]  \ . \label{replicatedeig}
\end{align}
In order to decouple sites, we introduce the functional order parameter

\begin{equation}
\varphi\left(\vec{v}\right)=\frac{1}{N}\sum_{i=1}^{N}\prod_{a=1}^{n}\delta\left(v_{a}-v_{ia}\right)\,,
\end{equation}
where the symbol $\vec{v}$ denotes a $n$-dimensional
vector in replica space. We enforce its definition using the integral identity

\begin{equation}
1=\int N\mathcal{D}\varphi\mathcal{D}\hat{\varphi}\exp\left\{ -\mathrm{i}\int\mathrm{d}\vec{v}\ \hat{\varphi}\left(\vec{v}\right)\left[N\varphi\left(\vec{v}\right)-\sum_{i}\prod_{a=1}^{n}\delta\left(v_{a}-v_{ia}\right)\right]\right\} \ .
\end{equation}
In terms of this order parameter and its conjugate, the replicated partition function can be written as

\begin{align}
\nonumber \left\langle Z^{n}\right\rangle _{J}  &=\left(\frac{\beta}{4\pi}\right)^n N\int\mathcal{D}\varphi\mathcal{D}\hat{\varphi}\mathrm{d}\vec{\lambda}\exp\left( -\mathrm{i}N\int\mathrm{d}\vec{v}\hat{\varphi}\left(\vec{v}\right)\varphi\left(\vec{v}\right)\right)\\
\nonumber & \times\exp\left[\frac{Nc}{2}\int\mathrm{d}\vec{v}\mathrm{d}\vec{v^\prime}\varphi(\vec{v})\varphi(\vec{v^\prime})\left(\left\langle \mathrm{e}^{\beta K\sum_{a}v_{a}v_{a}^{'}}\right\rangle _{K}-1\right)\right] \exp\left({ \mathrm{i}\frac{\beta}{2}N\sum_{a}\lambda_{a}} \right)\\
 & \times\int\prod_{a=1}^{n}\mathrm{d}\bm{v}_{a}\exp\left( -\mathrm{i}\frac{\beta}{2}\sum_{a}\sum_{i}\lambda_{a}v_{ia}^{2}\right) \exp\left[ \mathrm{i}\sum_{i}\int\mathrm{d}\vec{v}\hat{\varphi}\left(\vec{v}\right)\prod_{a=1}^{n}\delta\left(v_{a}-v_{ia}\right)\right] \ .
\end{align}
The multiple integral in the last line above factorises into $N$ identical copies of the same $n$-dimensional integral, and can thus be written as

\begin{equation}
I=\exp \left[N\mathrm{Log}\int\mathrm{d}\vec{v}\exp\left( -\mathrm{i}\frac{\beta}{2}\sum_{a}\lambda_{a}v_{a}^{2}+\mathrm{i}\hat{\varphi}(\vec{v})\right)\right] \ ,
\end{equation}
where $\mathrm{Log}$ denotes the principal branch of the complex logarithm.

Therefore, the replicated partition function takes a form amenable to a saddle point evaluation for large $N$ (where we assume we can safely exchange the limits $n\to 0$ and $N\to\infty$)

\begin{equation}
\left\langle Z^{n}\right\rangle _{J}\propto\int\mathcal{D}\varphi\mathcal{D}\hat{\varphi}\mathrm{d}\vec{\lambda}\exp\left(NS_{n}[\varphi,\hat{\varphi},\vec{\lambda}]\right)\ ,
\end{equation}
where
\begin{equation}
S_{n}[\varphi,\hat{\varphi},\vec{\lambda}]=S_{1}\left[\varphi,\hat{\varphi}\right]+S_{2}\left[\varphi\right]+S_{3}(\vec{\lambda})+S_{4}[\hat{\varphi},\vec{\lambda}]\ ,
\end{equation}
and

\begin{align}
S_{1}[\varphi,\hat{\varphi}] & =-\mathrm{i}\int\mathrm{d}\vec{v}\hat{\varphi}(\vec{v})\varphi(\vec{v})\ ,\label{eq:S1}\\
S_{2}[\varphi] & =\frac{c}{2}\int\mathrm{d}\vec{v}\mathrm{\mathrm{d}}\vec{v^\prime}\varphi(\vec{v})\varphi(\vec{v^\prime})\left(\left\langle \mathrm{e}^{\beta K\sum_{a}v_{a}v_{a}^{'}}\right\rangle _{K}-1\right)\ ,\label{eq:S2}\\
S_{3}(\vec{\lambda}) & =\mathrm{i}\frac{\beta}{2}\sum_{a}\lambda_{a}\ ,\label{eq:S3}\\
S_{4}[\hat{\varphi},\vec{\lambda}] & =\mathrm{Log}\int\mathrm{\mathrm{d}}\vec{v}\exp\left[ -\mathrm{i}\frac{\beta}{2}\sum_{a}\lambda_{a}v_{a}^{2}+\mathrm{i}\hat{\varphi}(\vec{v})\right] \ .
\end{align}

The stationarity of the action $S_{n}$ w.r.t. variations of $\varphi$
and $\hat{\varphi}$ requires that the order parameter at the saddle point $\varphi^\star$ and its
conjugate $\hat{\varphi}^\star$ satisfy the following coupled equations

\begin{align}
\mathrm{i}\hat{\varphi}^\star(\vec{v}) &=c\int\mathrm{d}\vec{v^\prime}\varphi^\star(\vec{v^\prime})\left[ \left\langle \exp\left(\beta K\sum_a v_{a}v_{a}^\prime\right)\right\rangle _{K}-1\right]\ ,\label{eq:rho_stat}\\
\varphi^\star(\vec{v}) &=\frac{\exp\left[ -\mathrm{i}\frac{\beta}{2}\sum_{a}\lambda_{a}v_{a}^{2}+\mathrm{i}\hat{\varphi}^\star\left(\vec{v}\right)\right] }{\int\mathrm{d}\vec{v^\prime}\exp\left[ -\mathrm{i}\frac{\beta}{2}\sum_{a}\lambda_{a}v_{a}^{\prime 2}+\mathrm{i}\hat{\varphi}^\star(\vec{v^\prime})\right] }\ ,\label{eq:hat_rho_stat}
\end{align}
which have to be solved together with the stationarity conditions w.r.t each component $\lambda_{\bar{a}}$ of $\vec{\lambda}$ 

\begin{equation}
1=\frac{\int\mathrm{d}\vec{v}\exp\left[ -\mathrm{i}\frac{\beta}{2}\sum_{a}\lambda_{a}v_{a}^{2}+\mathrm{i}\hat{\varphi}^\star(\vec{v})\right] v_{\bar{a}}^{2}}{\int\mathrm{d}\vec{v}\exp\left[ -\mathrm{i}\frac{\beta}{2}\sum_{a}\lambda_{a}v_{a}^{2}+\mathrm{i}\hat{\varphi}^\star\left(\vec{v}\right)\right] }\qquad\forall\bar{a}=1,\ldots,n\ .\label{eq:lambda_stat}
\end{equation}

The equations \eqref{eq:rho_stat} and \eqref{eq:hat_rho_stat} bear a striking resemblance with the saddle-point equations leading to the spectral density of Erd\H{o}s-R\'enyi random graphs \cite{Rodgers1988,Kuehn2008}, except for the fact that the ``Hamiltonian"  of our problem is real-valued and includes the inverse temperature $\beta$. Following \cite{Kuehn2008}, we will now search for replica-symmetric solutions written in the form of superpositions of uncountably infinite Gaussians with a non-zero mean. This ansatz will be preserving permutational symmetry between replicas, but (at odds with the choice in \cite{Kuehn2008}) not the rotational invariance in the space of replicas\footnote{A rotationally invariant ansatz would not produce a physically meaningful result for this problem.}:

 \begin{align}
\lambda_{\bar{a}} &=\lambda\qquad\forall \bar{a}=1,\ldots,n\ ,\\
\varphi^\star(\vec{v}) &=\int\mathrm{d}\omega\mathrm{d}h\ \pi\left(\omega,h\right)\prod_{a=1}^{n}\frac{1}{Z_\beta(\omega,h)}\exp\left[ -\frac{\beta}{2}\omega v_{a}^{2}+\beta hv_{a}\right] \ ,\label{eq:ansatz_rep_1}\\
\mathrm{i}\hat{\varphi}^\star(\vec{v}) &=\hat{c}\int\mathrm{d}\hat{\omega}\mathrm{d}\hat{h}\ \hat{\pi}(\hat{\omega},\hat{h})\prod_{a=1}^{n}\exp\left[ \frac{\beta}{2}\hat{\omega}v_{a}^{2}+\beta\hat{h}v_{a}\right] \ ,\label{eq:ansatz_rep_2}
\end{align}
where 

\begin{equation}
Z_\beta(x,y)=\sqrt{\frac{2\pi}{\beta x}}\exp\left(\frac{\beta y^{2}}{2 x}\right)\ .\label{eq:zomega_er}
\end{equation}

To justify the procedure above, on one hand the replica symmetric ansatz has been known for quite a while to lead to the correct results for the spectral problem of sparse random matrices \cite{Edwards1976,Rodgers1988,Kuehn2008,Khorunzhy2004}. On the other hand, it is known that expressing the order parameter as a superposition of Gaussian pdfs provides the correct solution for harmonically coupled system \cite{Kuehn2007}.

In \eqref{eq:ansatz_rep_1} and \eqref{eq:ansatz_rep_2}, $\pi$ and $\hat\pi$ are normalised joint pdfs of the parameters appearing in the Gaussian distributions, while $\hat{c}$ is introduced taking into account that $\mathrm{i}\hat\varphi(\vec{v})$ needs not be normalised. The advantage of writing an ansatz in this form is that - once inserted into \eqref{eq:rho_stat} and \eqref{eq:hat_rho_stat} - it makes it possible to perform explicitly the $\vec{v}$-integrals, eventually leading to simpler coupled equations for $\pi$ and $\hat\pi$, as detailed below. The convergence of the $\vec{v}$-integrals will also impose the following conditions on $\omega$ and $\hat\omega$: $\omega>\hat\omega$ and $\omega >\zeta$ (where $\zeta$ is the upper edge of the support of the pdf $p(K)$ of bond weights).

As a further remark, the different signs in front of $\omega$ and $\hat\omega$ in \eqref{eq:ansatz_rep_1} and \eqref{eq:ansatz_rep_2} are picked with an eye towards performing the subsequent $\vec{v}$-integrals: since $\mathrm{i}\hat{\varphi}^\star(\vec{v})$ is not a pdf, $\hat\omega$ being positive is not problematic.

Rewriting the action in terms of $\pi$ and $\hat\pi$, after performing the $\vec{v}$-integrations, and extracting the leading $n\to 0$ contribution yields

\begin{align}
S_{1}[\pi,\hat{\pi}] & =-\hat{c}-\hat{c}n\int\mathrm{d}\pi(\omega,h)\mathrm{d}\hat{\pi}(\hat{\omega},\hat{h})\ln\frac{Z_\beta(\omega-\hat{\omega},h+\hat{h})}{Z_\beta(\omega,h)}\ ,\label{eq:S1_pi} & {}\\
S_{2}[\pi] & =\frac{c}{2}n\int\mathrm{d}\pi(\omega,h)\mathrm{d}\pi(\omega',h')\left\langle \ln\frac{Z^{(2)}_\beta\left(\omega,\omega',h,h',K\right)}{Z_\beta\left(\omega,h\right)Z_\beta\left(\omega',h'\right)}\right\rangle _{K}\ ,\label{eq:S2_pi}\\
S_{3}(\lambda) & =\mathrm{i}\frac{\beta}{2}n\lambda\ ,\label{eq:S3_pi}\\
S_{4}[\hat{\pi},\lambda] & =\hat{c}+n\sum_{s=0}^{\infty}p_{\hat{c}}\left(s\right)\int\{ \mathrm{d}\hat{\pi}\} _{s}~\mathrm{Log}~Z_\beta\left(\mathrm{i}\lambda-\{ \hat{\omega}\} _{s},\{ \hat{h}\} _{s}\right)\label{S4beforesaddle}\ ,
\end{align}
where we have introduced the shorthands 
\begin{equation}
Z^{(2)}_\beta(\omega,\omega',h,h',K)=Z_\beta(\omega',h')Z_\beta\left(\omega-\frac{K^{2}}{\omega'},h+\frac{h'K}{\omega'}\right)
\end{equation}
and $\{\mathrm{d}\hat{\pi}\} _{s}=\prod_{\ell=1}^{s}\mathrm{d}\hat{\omega}_{\ell}\mathrm{d}\hat{h}_{\ell}\hat{\pi}(\hat{\omega}_{\ell},\hat{h}_{\ell})$,
along with $\{ \hat{\omega}\} _{s}=\sum_{\ell=1}^{s}\hat{\omega}_{\ell}$
and $\{ \hat{h}\} _{s}=\sum_{\ell=1}^{s}\hat{h}_{\ell}$.
The symbol $p_{\hat{c}}(s)$ denotes a Poissonian degree
distribution 
$
p_{\hat{c}}(s)=  \hat{c}^{s} \mathrm{e}^{-\hat{c}}/s!
$
with mean $\hat{c}$, which naturally arises in the calculation.
We note that the $\mathcal{O}(1)$ terms in $S_{1}$ and $S_{4}$
cancel, so that $S_{n}=\mathcal{O}(n)$ as expected.

The full action in terms of $\pi$ and $\hat{\pi}$
now reads
\begin{equation}
S_n=S_{1}[\pi,\hat{\pi}]+S_{2}[\pi]+S_{3}(\lambda)+S_{4}[\hat{\pi},\lambda]\ .\label{eq:action_pihatpi}
\end{equation}
The stationarity condition w.r.t $\lambda$ entails

\begin{equation}
\frac{\partial S}{\partial\lambda}\Big|_{\lambda=\lambda^\star}=0\Rightarrow 1=\sum_{s=0}^{\infty}p_{\hat{c}}(s)\int\{\mathrm{d}\hat{\pi}\} _{s}\langle v^{2}\rangle_{\bar P} \ ,\label{eq:stat_k}
\end{equation}
where the average $\langle \cdot \rangle_{\bar{P}} $ is taken with respect to
the Gaussian measure
\begin{equation}
\bar{P}(v) = \sqrt{\frac{\beta\left(\mathrm{i}\lambda^\star-\{ \hat{\omega}\} _{s}\right)}{2\pi}}\exp\left[ -\frac{\beta}{2}\left(\mathrm{i}\lambda^\star-\{ \hat{\omega}\} _{s}\right)\left(v-\frac{\{ \hat{h}\} _{s}}{\mathrm{i}\lambda^\star-\{ \hat{\omega}\} _{s}}\right)^{2}\right]\ .\label{eq:p_bar}
\end{equation}
More explicitly, \eqref{eq:stat_k} reads
\begin{equation}
1=\sum_{s=0}^{\infty}p_{\hat{c}}(s)\int\{\mathrm{d}\hat{\pi}\} _{s}\left[\frac{1}{\beta(\mathrm{i}\lambda^\star-\{ \hat{\omega}\} _{s})}+\left(\frac{\{ \hat{h}\} _{s}}{\mathrm{i}\lambda^\star-\{ \hat{\omega}\} _{s}}\right)^{2}\right].\label{lambdastarcondition}
\end{equation}
We note that the $\beta$-dependent term vanishes as $\beta\rightarrow\infty$.

The stationarity condition with respect to variations of $\pi$, 
$
\frac{\delta S}{\delta\pi}  =  0
$, 
entails the condition
\begin{equation}
\frac{\hat{c}}{c}\int\mathrm{d}\hat{\pi}(\hat{\omega},\hat{h})\ln\frac{Z_\beta(\omega-\hat{\omega},h+\hat{h})}{Z_\beta(\omega,h)}=\int\mathrm{d}\pi(\omega',h')\left\langle \ln\frac{Z^{(2)}_\beta(\omega,\omega',h,h',K)}{Z_\beta(\omega,h)}\right\rangle _{K}+\frac{\gamma}{c}\ ,\label{eq:pi_variation}
\end{equation}
where $\gamma$ is a Lagrange multiplier
introduced to enforce the normalisation of $\pi$. Given the definition of $Z^{(2)}_\beta$, \eqref{eq:pi_variation} is equivalent to 
\begin{equation}
\frac{\hat{c}}{c}\int\mathrm{d}\hat{\pi}(\hat{\omega},\hat{h})\ln Z_\beta(\omega-\hat{\omega},h+\hat{h})=\int\mathrm{d}\pi(\omega',h')\left\langle \ln Z_\beta\left(\omega-\frac{K^{2}}{\omega'},h+\frac{h'K}{\omega'}\right)\right\rangle _{K}+\frac{\gamma}{c}\ .\label{eq:stat_pi}
\end{equation}
The condition that \eqref{eq:stat_pi} must hold for all $\omega$ and $h$ can be translated into
\begin{equation}
\hat{\pi}(\hat{\omega},\hat{h})=\int\mathrm{d}\omega\mathrm{d}h~\pi(\omega,h)\left\langle \delta\left(\hat{\omega}-\frac{K^{2}}{\omega}\right)\delta\left(\hat{h}-\frac{hK}{\omega}\right)\right\rangle _{K}\ ,\label{eq:pi_hat}
\end{equation}
where $c=\hat{c}$ to enforce normalization of $\hat{\pi}$.

Similarly, the stationarity condition with respect to variations of $\hat{\pi}$, 
$
\frac{\delta S}{\delta\hat{\pi}}=0,
$
produces the condition

\begin{align}
\nonumber\int\mathrm{d}\pi(\omega,h)\ln Z_\beta\left(\omega-\hat{\omega},h+\hat{h}\right) &=\sum_{s=1}^{\infty}\frac{s}{c}p_{c}(s)\int\{\mathrm{d}\hat{\pi}\} _{s-1}\mathrm{Log}~ Z_\beta(\mathrm{i}\lambda^\star-\{ \hat{\omega}\} _{s-1}-\hat{\omega},\{ \hat{h}\} _{s-1}+\hat{h})\\
&+\frac{\hat\gamma}{c}\ ,\label{eq:stat_pihat2}
\end{align}
where $\hat\gamma$ is the Lagrange multiplier enforcing the normalisation of $\hat\pi$.
We can then conclude that the saddle-point pdf $\pi$ must satisfy
\begin{equation}
\pi(\omega,h)=\sum_{s=1}^{\infty}\frac{s}{c}p_{c}(s)\int\{ \mathrm{d}\hat{\pi}\} _{s-1}\delta\left(\omega-(\mathrm{i}\lambda^\star-\{ \hat{\omega}\} _{s-1})\right)\delta(h-\{ \hat{h}\} _{s-1})\ .\label{eq:pi_firstform}
\end{equation}
Inserting \eqref{eq:pi_hat} into \eqref{eq:pi_firstform} yields, after simple algebra
\begin{equation}
\pi(\omega,h)=\sum_{s=1}^{\infty}\frac{s}{c}p_{c}(s)\int\{\mathrm{d}\pi\} _{s-1}\left\langle \delta\left(\omega-\left(\mathrm{i}\lambda^\star-\sum_{\ell=1}^{s-1}\frac{K_{\ell}^{2}}{\omega_{\ell}}\right)\right)\delta\left(h-\sum_{\ell=1}^{s-1}\frac{h_{\ell}K_{\ell}}{\omega_{\ell}}\right)\right\rangle _{\{ K\}_{s-1} }\ ,\label{eq:pi}
\end{equation}
where the brackets $\left\langle\cdot \right\rangle _{\{ K\}_{s-1} }$
denote averaging with respect to a collection of $s-1$ i.i.d. random variables $K_{\ell}$, each drawn from the bond weight pdf $p(K)$. 

We recall at this point that the replica derivation started under the simplifying assumption that the connectivity distribution was that of a standard Erd\H{o}s-R\'enyi graph (see \eqref{eq:joint_er}). This implies that the degree distribution $p_c(s)$ - naturally appearing in \eqref{eq:pi} - is a Poisson distribution with unbounded support. However, Eq. \eqref{eq:pi} remains formally valid for \emph{any} degree distribution $p_c(s)$ with finite mean $c$. In our case, it is then necessary to consider \eqref{eq:degree_pmf} and manually correct\footnote{Obviously, the ``truncated" Eq. \eqref{eq:pi_trunc} would have been obtained anyway without any shortcuts, had we started from the exact connectivity distribution \eqref{eq:boundedER_connectivity} at the outset. This is explicitly shown in \ref{sec:No_shortcuts}.} \eqref{eq:pi} to account for the existence of a maximal degree, therefore yielding
\begin{equation}
\pi(\omega,h)=\sum_{s=1}^{k_{\mathrm{max}}}r(s)\int\{\mathrm{d}\pi\} _{s-1}\left\langle \delta\left(\omega-\left(\mathrm{i}\lambda^\star-\sum_{\ell=1}^{s-1}\frac{K_{\ell}^{2}}{\omega_{\ell}}\right)\right)\delta\left(h-\sum_{\ell=1}^{s-1}\frac{h_{\ell}K_{\ell}}{\omega_{\ell}}\right)\right\rangle _{\{ K\}_{s-1} }\ ,\label{eq:pi_trunc}
\end{equation}
where $r(s)$ is the link-degree distribution \eqref{eq:pmf_link}. Note that \eqref{eq:pi_trunc} is formally identical to the self-consistent equation \eqref{eq:q} found for the cavity field pdf, after the identification $\pi(\omega,h)\equiv q(\omega,h)$.

The constant term $\lambda\equiv \mathrm{i}\lambda^\star$ -- which turns out to be real-valued -- needs to be tuned so as to enforce \eqref{lambdastarcondition} for $\beta\to\infty$, which reads (trading $\hat\pi$ for $\pi$)

\begin{equation}
1=\sum_{s=0}^{k_{\mathrm{max}}}p_c(s)\int\{\mathrm{d}\pi\} _{s}\left\langle\left(\frac{\sum_{\ell=1}^s \frac{h_\ell K_\ell}{\omega_\ell}}{\lambda-\sum_{\ell=1}^s \frac{K_\ell^2}{\omega_\ell}}\right)^{2}\right\rangle_{\{K\}_s}\ ,\label{eq:lambdastarconditionV2}
\end{equation}
where -- to avoid introducing more cumbersome notations -- $p_c(s)$ now indicates the actual bounded degree distribution \eqref{eq:degree_pmf}.

Surprisingly, even though the cavity and replica methods
depart from completely different assumptions, they converge towards
the same result: this has been already shown in \cite{Slanina2011}
for the spectral problem in the Erd\H{o}s-R\'enyi case. 

A few remarks are in order: 
\begin{itemize}
\item For the action to converge, we have obtained the following conditions $\omega>\zeta$, $\omega>\hat\omega$ and $\lambda\equiv\mathrm{i}\lambda^\star>\{\hat\omega\}_s$,
where $\zeta$ is the upper bound of the support of the bond weights $p(K)$.
\item Thanks to the structure of $\hat\pi$ \eqref{eq:pi_hat}, the entire action can be just expressed in term of $\pi$ \eqref{eq:pi} and $\lambda$ \eqref{eq:lambdastarconditionV2}.
\item The value of $\lambda\equiv \mathrm{i}\lambda^\star$ is real, and corresponds to the typical value of the largest eigenvalue $\langle\lambda_1\rangle_J$, as will be shown in subsection \ref{sec:ERaction}. This is of course again compatible with the cavity results.
\item In Eq. \eqref{eq:pi_trunc}, the contribution corresponding to $s=1$ in the sum gives rise to the term $\delta(\omega-\lambda)$ on the right hand side. Therefore, we expect to see a pronounced peak at the location of $\lambda=\langle\lambda_1\rangle_J$ in the plot of the marginal pdf $\pi(\omega)=\int\mathrm{d}h~\pi(\omega,h)$, once the contributions coming from nodes of different degrees are ``unpacked". This is confirmed in Fig. \ref{fig:weightedER} below.
\item Both the cavity and replica approaches can be safely extended to non-Poissonian degree distributions as well, as long as the mean connectivity $c$ remains finite as $N\rightarrow\infty$,
thus considerably enlarging the class of models for which the equivalence between cavity and replicas holds true.
\end{itemize}

\subsubsection{Erd\H{o}s-R\'enyi graph: weighted adjacency matrix.\label{sec:ERaction}}

We proceed here with the case of a weighted adjacency matrix of sparse Erd\H{o}s-R\'enyi graphs, with bounded maximal degree and bond weights drawn from the pdf $p(K)$. The pure $\{0,1\}$-adjacency matrix case is recovered considering $p(K)=\delta(K-1)$.
Given the distributions \eqref{eq:pi_trunc} and \eqref{eq:pi_hat} at stationarity
and recalling \eqref{eq:zomega_er}, the $\mathcal{O}(n)$ terms of the action
$S_{n}$ in \eqref{eq:action_pihatpi} - keeping only the leading $\beta\rightarrow\infty$ term - are expressed
as:

\begin{align}
\nonumber S_{1}\left[\pi,\hat{\pi}\right]  &=-nc\int\mathrm{d}\pi(\omega,h)\mathrm{d}\hat{\pi}(\hat\omega,\hat h)\ln\frac{Z_\beta(\omega-\hat{\omega},h+\hat{h})}{Z_\beta(\omega,h)} \\
 &\simeq -nc\frac{\beta}{2}\int\mathrm{d}\pi(\omega,h)\mathrm{d}\pi(\omega',h')\left\langle\frac{\left(h+\frac{h'K}{\omega'}\right)^{2}}{\omega-\frac{K^2}{\omega'}}-\frac{h^2}{\omega}\right\rangle_K =-nc\frac{\beta}{2}I_{1}\ ,
\end{align}

\begin{align}
\nonumber S_{2}[\pi] & =n\frac{c}{2}\int\mathrm{d}\pi(\omega,h)\mathrm{d}\pi(\omega',h')\left\langle\ln\frac{Z_\beta\left(\omega-\frac{K^2}{\omega'},h+\frac{h'K}{\omega'}\right)}{Z_\beta(\omega,h)}\right\rangle_K\\
 & \simeq n c\frac{\beta}{4}\int\mathrm{d}\pi(\omega,h)\mathrm{d}\pi(\omega',h')\left\langle\frac{\left(h+\frac{h^{'}K}{\omega'}\right)^{2}}{\omega-\frac{K^2}{\omega^{'}}}-\frac{h^{2}}{\omega}\right\rangle_K=nc\frac{\beta}{4}I_1\ ,
\end{align}

\begin{align}
S_{3}\left(\lambda\right) & =\frac{\beta}{2}n\lambda\ ,
\end{align}

\begin{align}
S_{4}[\hat{\pi},\lambda] & =n\sum_{s=0}^{\infty}p_{c}(s)\int\left[\prod_{\ell=1}^{s}\mathrm{d}\hat{\pi}(\hat{\omega}_{\ell},\hat{h}_{\ell})\right]\mathrm{Log}~Z_\beta\left(\lambda-\{ \hat{\omega}\} _{s},\{ \hat{h}\} _{s}\right) & {}\nonumber \\
 & \simeq n\frac{\beta}{2}\sum_{s=0}^{\infty}p_{c}(s)\int\left[\prod_{\ell=1}^{s}\mathrm{d}\hat{\pi}(\hat{\omega}_{\ell},\hat{h}_{\ell})\right]\left(\frac{\left(\sum_{\ell=1}^{s}\hat{h}_{\ell}\right)^{2}}{\lambda-\sum_{\ell=1}^{s}\hat{\omega}_{\ell}}\right)\label{S4afterbeta}\ .
\end{align}
Multiplying and dividing the integrand of \eqref{S4afterbeta} by $\lambda-\sum_{\ell=1}^s\hat\omega_\ell$, and using \eqref{lambdastarcondition} (for $\beta\to\infty$), we get
\begin{align}
S_{4}[\hat{\pi},\lambda] &=n\frac{\beta}{2}\lambda-n\frac{\beta}{2}\sum_{s=1}^{\infty}p_{c}(s)s\int\mathrm{d}\hat{\pi}(\hat{\omega},\hat{h})\{\mathrm{d}\hat\pi\}_{s-1}\left(\frac{\sum_{\ell=1}^{s-1}\hat{h}_{\ell}+\hat h}{\lambda-\sum_{\ell=1}^{s-1}\hat{\omega}_{\ell}-\hat\omega}\right)^2\hat\omega\ .
\end{align}
Multiplying the second term by $1=\int\mathrm{d}\omega\mathrm{d}h\delta\left(\omega-(\lambda-\{ \hat{\omega}\} _{s-1})\right)\delta(h-\{ \hat{h}\} _{s-1})$, and using \eqref{eq:pi_firstform}, we obtain (after some manipulations)
\begin{align}
S_{4}[\pi,\lambda] &=n\frac{\beta}{2}\lambda-nc\frac{\beta}{2}\int \mathrm{d}\pi(\omega,h)\mathrm{d}\pi(\omega',h')\left\langle\frac{K^2}{\omega'}\left(\frac{h+h'K/\omega'}{\omega-K^2/\omega'}\right)^2\right\rangle_K=n\frac{\beta}{2}\lambda-nc\frac{\beta}{2}I_2\ .
\end{align}
Summing up all terms, the action at the saddle point reads 

\begin{equation}
S_{n}=\frac{n\beta}{2}\left(-\frac{c}{2}I_1-c I_2+2\lambda\right)\ ,
\end{equation}
which would imply from \eqref{formulalargest} for the average of the largest eigenvalue the formula
\begin{equation}
\left\langle \lambda_{1}\right\rangle _{J}=-\frac{c}{2}I_1-c I_2+2\lambda\ .
\end{equation}
However, we were able to numerically show that at the saddle point
\begin{equation}
\lambda=c\left(I_2+\frac{1}{2}I_1\right)\ ,
\label{eq:ilambda_integral}
\end{equation} 
implying that \begin{equation}
\left\langle \lambda_{1}\right\rangle _{J}=\lambda\ ,
\end{equation} 
as expected from the corresponding cavity calculation. The identity \eqref{eq:ilambda_integral} can be more easily checked numerically once expressed  in the alternative way

\begin{equation}
\left\langle \lambda_{1}\right\rangle _{J}=\lambda=c\int\mathrm{d}\pi(\omega,h)\mathrm{d}\pi(\omega',h')\left\langle\left(\frac{h+\frac{h'K}{\omega'}}{\omega-\frac{K^2}{\omega'}}\right)\left(\frac{h'+\frac{hK}{\omega}}{\omega'-\frac{K^2}{\omega}}\right)K\right\rangle_K ,\label{eq:ilambdastar}
\end{equation}
which has the additional advantage of showing explicitly that $\lambda\equiv\mathrm{i}\lambda^\star$ is indeed a real-valued quantity.

\begin{figure}
\begin{centering}
\includegraphics[scale=0.4]{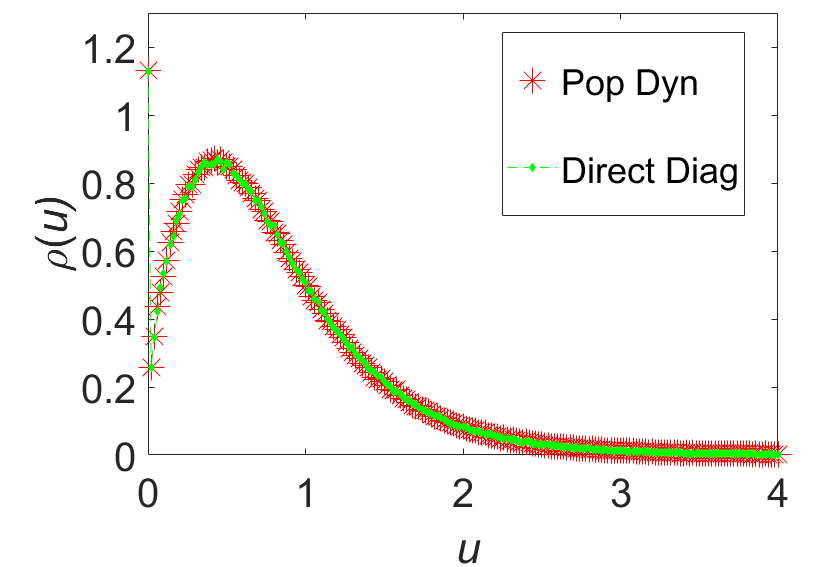}\includegraphics[scale=0.4]{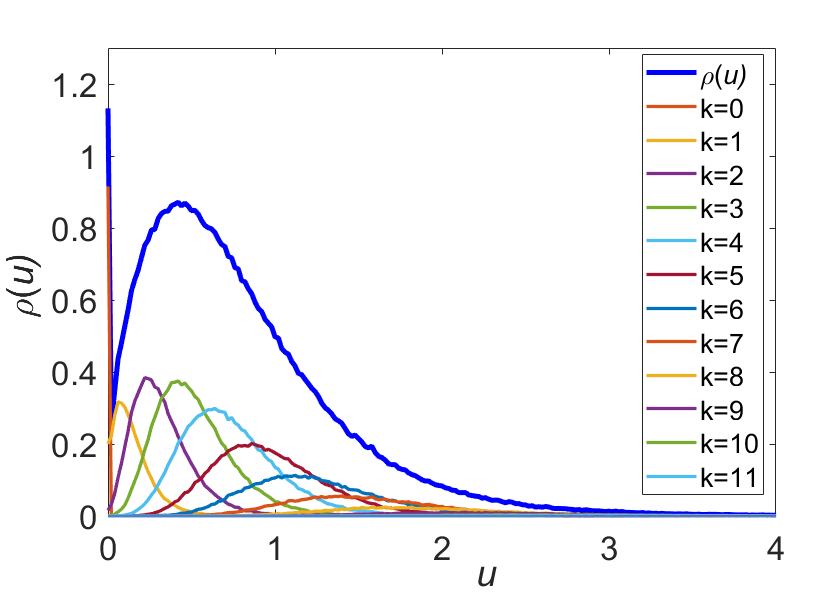}
\par\end{centering}
\begin{centering}
\includegraphics[scale=0.4]{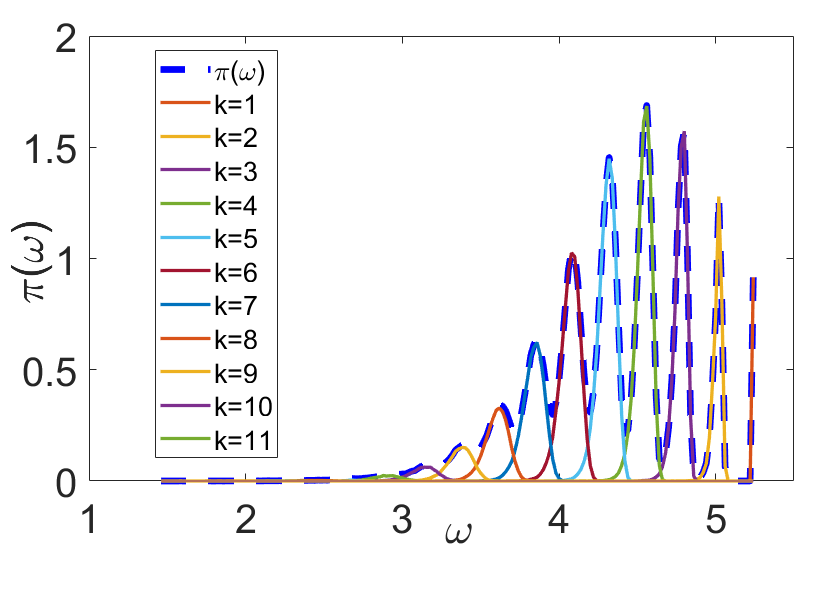}\includegraphics[scale=0.4]{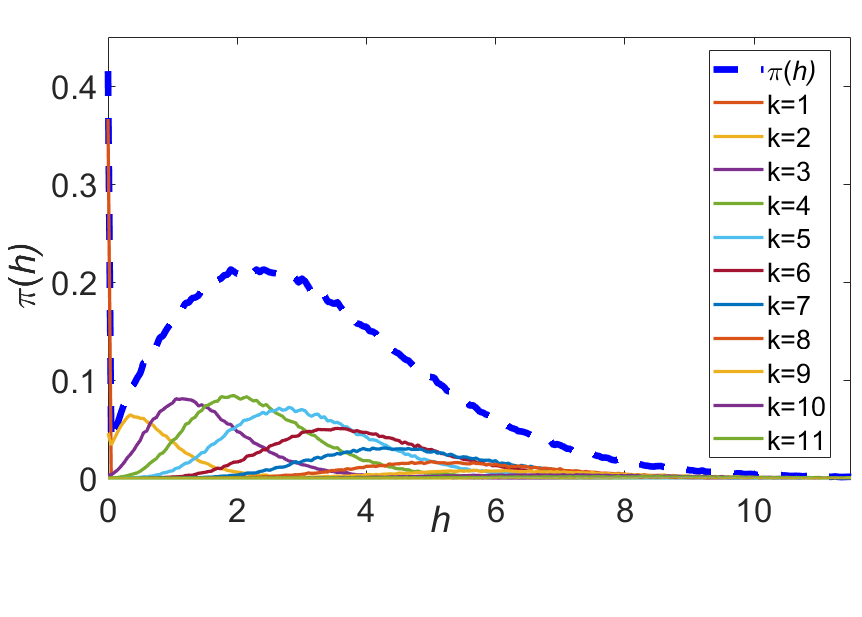}
\par\end{centering}
\begin{centering}
\caption{All panels refer to the Erd\H{o}s-R\'enyi adjacency matrix in the limit $N\rightarrow\infty$. The plots are obtained via the population dynamics algorithm described in Section \ref{sec:Population-Dynamics}. In all cases, the mean connectivity is $c=4$, $k_{\mathrm{max}}=16$ and the population size is $N_{P}=10^{6}$. The resulting typical top eigenvalue is $\left\langle \lambda_{1}\right\rangle_J\approx 5.254$. {\bf Top left panel}: comparison between results
for the density of top eigenvector's components (see \eqref{eq:first_eigv_comp_denisity_cavity}
or equivalently \eqref{eq:density_with_pi}),
obtained with population dynamics (red stars) and direct diagonalisation (green diamonds). {\bf Top right panel}: density
of the top eigenvector's components in the Erd\H{o}s-R\'enyi case: the thick blue
line is the full pdf, whereas the thinner curves underneath
indicate the contributions from nodes of various degree from $k=0$ to $k=16$. Only the degree contributions up to $k=11$ are labelled: all the other (larger) degree contributions are barely distinguishable as they fall on top of each other in the tail of the distribution. {\bf Bottom left
panel}: marginal distribution of the inverse single site variances $\omega$. The thick dashed line represents the full pdf, the thinner curves underneath stand for the single degree contributions, from $k=1$ to $k=16$. The rightmost peak at $\omega=\lambda$ corresponds to $k=1$: the degree decreases as the peaks are centered at lower $\omega$. Also in this case, only the degree contributions up to $k=11$ are highlighted. {\bf Bottom right panel}: marginal pdf
of the single-site bias fields $h$. Again, the thick dashed line represents the full distribution,
the thinner curves stand for the degree contributions from $k=1$ to $k=16$. The leftmost peak at $h=0$ corresponds to $k=1$: as $h$ grows, the pdf $\pi(h)$ receives contributions from higher degrees. Also in this case, only the degree contributions up to $k=11$ are highlighted.  }\label{fig:summary}
\par\end{centering}
\centering{}
\end{figure}

\begin{figure}
\begin{centering}
\includegraphics[scale=0.4]{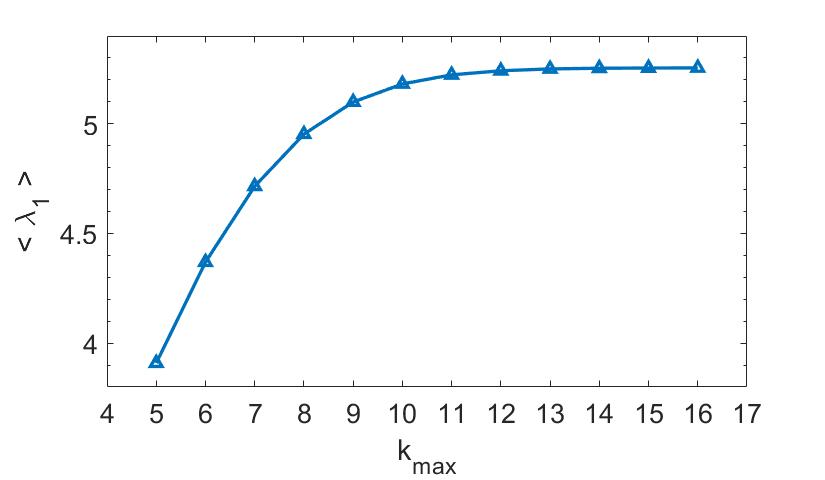}
\par\end{centering}
\begin{centering}
\caption{This panel refers to the behaviour of the typical largest eigenvalue in the Erd\H{o}s-R\'enyi adjacency matrix case as the maximum degree $k_{\mathrm{max}}$ is varied. The value of $\left\langle \lambda_1 \right\rangle$ is found via population dynamics for any fixed value of $k_{\mathrm{max}}$. Each value has been then checked against direct diagonalisation extrapolation at $N\to\infty$. The mean connectivity parameter $\bar{c}$ appearing in \eqref{eq:degree_pmf} is set to 4, whereas the population size is $N_{P}=10^6$ for any data point. Clearly, the mean degree $c$ tends to $\bar{c}=4$ as $k_{\mathrm{max}}$ increases.
As expected, $\left\langle \lambda_1 \right\rangle$ grows as $k_{\mathrm{max}}$ increases, but the growth becomes slower as the probability of finding a node of higher and higher degree becomes negligible even in the thermodynamic limit. 
}\label{Fig:lambda_vs_kmax}
\par\end{centering}
\end{figure}

\begin{figure}
\begin{centering}
\includegraphics[scale=0.45]{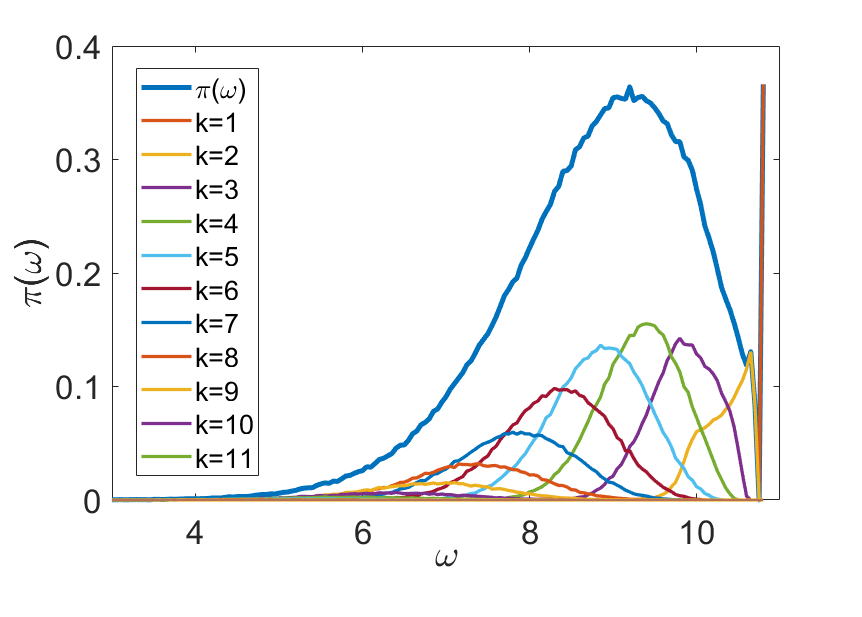}
\includegraphics[scale=0.45]{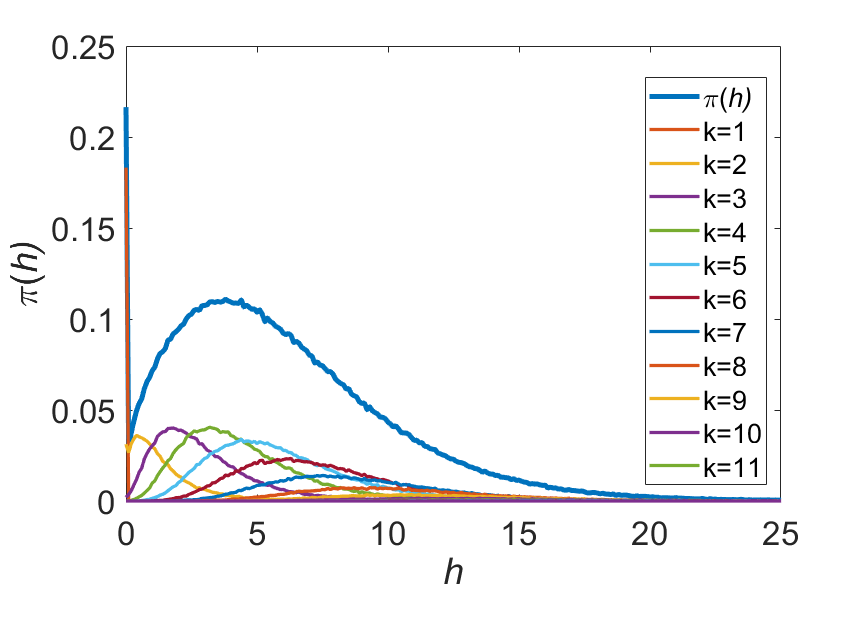}
\par\end{centering}

\begin{centering}
\caption{Marginal distributions $\pi(\omega)$ and $\pi(h)$ for Erd\H{o}s-R\'enyi weighted adjacency matrices  in the limit $N\rightarrow\infty$. The graphs are obtained via the population dynamics algorithm. Here the mean connectivity is $c=4$, $k_{\mathrm{max}}=16$ and the population size is $N_{P}=10^{6}$. The bond weight distribution is chosen to be uniform, specifically $p(K)=1/2$ for all $K\in[1,3]$. The resulting typical top eigenvalue is $\left\langle \lambda_{1}\right\rangle_J\approx 10.8407$. {\bf Top panel}: marginal distribution of the $\omega$-variables; the thick blue
line represents the full distribution, the thinner curves underneath correspond to the various degree contributions from $k=1$ up to $k=16$. The contribution of nodes with degree $k=1$ corresponds to the peak located at $\omega=\left\langle \lambda_{1}\right\rangle_J\approx 10.8407$, as expected from Eq. \eqref{eq:pi_trunc}. The peculiar structure of the distribution $\pi(\omega)$ in the case of the pure adjacency matrix (see Fig. \ref{fig:summary}) where every single degree contribution corresponds to a specific peak in $\pi(\omega)$ is lost here, due to the presence of nontrivial bond weights. As in Fig.  \ref{fig:summary}, only the degree contributions up to $k=11$ are labelled.
{\bf Bottom panel}: marginal distribution of the bias fields $h$; again, the thick blue line represents the full distribution, while the thinner curves underneath correspond to the different degree contributions. Again, only the degree contributions up to $k=11$ are labelled.}
\label{fig:weightedER}
\par\end{centering}
\centering{}
\end{figure}

The bottom panels in Fig. \ref{fig:summary} show the marginal distributions $\pi(\omega)=\int\de h~\pi(\omega,h)$ and $\pi(h)=\int\de\omega~\pi(\omega,h)$ for the case of a pure Erd\H{o}s-R\'{e}nyi $\{0,1\}$-adjacency matrix, for which $p(K)=\delta(K-1)$. Figure \ref{fig:weightedER} instead shows $\pi(\omega)$ and $\pi(h)$ for the case of a weighted Erd\H{o}s-R\'enyi adjacency matrix, with a uniform bond pdf $p(K)=1/2$ for $K\in(1,3)$. In Fig. \ref{Fig:lambda_vs_kmax}, we plot the behaviour of the typical largest eigenvalue as the maximum degree $k_{\mathrm{max}}$ is varied.

\subsubsection{Random regular graph: adjacency matrix.\label{sec:rrg}}

We now consider the simpler and analytically tractable case of the random regular graph (RRG). A RRG with connectivity $c$ is characterized by the property that every node has exactly $c$ neighbours, or equivalently every row of its $\{0,1\}$-adjacency matrix has exactly $c$ nonzero entries. This implies that the largest eigenvalue of such matrix is $\left\langle\lambda_1\right\rangle_J=\lambda=c$ (deterministically), and its corresponding eigenvector has all identical components $\bm{v}_1=(1,1,...,1)^T$.

In this case, the Poissonian degree distribution featuring in \eqref{eq:pi}
can be safely replaced by $\delta_{s,c}$. Furthermore, if we consider the pure adjacency matrix case (i.e. with $p(K)=\delta(K-1)$), \eqref{eq:pi} and \eqref{eq:lambdastarconditionV2} become
\begin{align}
\pi(\omega,h) &=\int\{\mathrm{d}\pi\} _{c-1} \delta\left(\omega-\left(\lambda-\sum_{\ell=1}^{c-1}\frac{1}{\omega_{\ell}}\right)\right)\delta\left(h-\sum_{\ell=1}^{c-1}\frac{h_{\ell}}{\omega_{\ell}}\right) \ ,\label{eq:piRRG}\\
1 &=\int\{\mathrm{d}\pi\} _{c}\left(\frac{\sum_{\ell=1}^c \frac{h_\ell }{\omega_\ell}}{\lambda-\sum_{\ell=1}^c \frac{1}{\omega_\ell}}\right)^{2}\ ,
\end{align}
which can be exactly solved by the ansatz
\begin{equation}
\pi(\omega,h)=\delta(\omega-\bar\omega)\delta(h-\bar{h})\ ,
\end{equation}
leading to the following equations for the parameters $\bar\omega,\bar{h}$ and $\lambda$
\begin{align}
\bar\omega &=\lambda-\frac{c-1}{\bar\omega}\label{eq:omegabar}\ ,\\
\bar{h} &=(c-1)\frac{\bar{h}}{\bar\omega}\label{eq:hbar}\ ,\\
1 &=\left(\frac{c\bar{h}/\bar\omega}{\lambda-c/\bar\omega}\right)^2\ .\label{eq:norm_rrg}
\end{align}
Eq. \eqref{eq:hbar} entails that $\bar\omega=c-1$. Then, inserting this value in \eqref{eq:omegabar}, we find $\lambda=c$.

The value of $\bar h$ can then be found exploiting the normalization condition \eqref{eq:norm_rrg}, yielding $\bar h=c-2$.

The action at the saddle-point reads then
\begin{equation}
S_n=n\frac{\beta}{2}\frac{{\bar{h}}^2}{\bar\omega}\left[-\frac{\bar\omega+1}{\bar\omega-1}+\frac{2}{\bar\omega-1}+1\right]+n\frac{\beta}{2}c=n\frac{\beta}{2}c\ ,
\end{equation}
and therefore, the typical largest eigenvalue is
\begin{equation}
\left\langle \lambda_{1}\right\rangle_J=\lim_{\beta\rightarrow\infty}\frac{2}{\beta N}\lim_{n\rightarrow0}\frac{1}{n}Nn\frac{\beta}{2}c=c\ ,
\end{equation}
equal to $\lambda$ as expected.

\subsection{Density of the top eigenvector's components}\label{sec:replica_eigenvector}

In this statistical mechanics framework, the quantity

\begin{equation}
\tilde{\rho}_{\beta}\left(u\right)=\left\langle{\frac{1}{N}\sum_{i=1}^{N}\delta\left(u-v_{i}\right)}\right\rangle \label{eq:instance_density}
\end{equation}
is defined such that in the limit $\beta\rightarrow\infty$ it gives
the density of the top eigenvector components for a given $N\times N$
sparse symmetric random matrix $J$. The simple angle brackets $\left\langle ... \right\rangle$  stands for
thermal averaging, i.e. with respect to the Gibbs-Boltzmann
distribution \eqref{eq:hard} of the system
\begin{equation}
P_{\beta,J}(\bm{v})=\frac{\exp\left(\frac{\beta}{2}\left(\bm{v},J\bm{v}\right)\right) \delta\left(\left|\bm{v}\right|^{2}-N\right)}{\int\mathrm{d}\bm{v}'\exp\left( \frac{\beta}{2}(\bm{v}',J\bm{v}')\right) \delta\left(\left|\bm{v}'\right|^{2}-N\right)}\ .
\end{equation}
Defining an auxiliary partition function as 

\begin{equation}
Z^{(\beta)}_\epsilon(t,J;u)=\int\mathrm{d}\bm{v}\exp\left[\frac{\beta}{2}\left(\bm{v},J\bm{v}\right)+\beta t\sum_{i}\delta_\epsilon\left(u-v_{i}\right)\right] \delta\left(\left|\bm{v}\right|^{2}-N\right)\ ,
\end{equation}
where $\delta_\epsilon$ is a smooth regulariser of the delta function, the quantity \eqref{eq:instance_density} can be formally expressed as 

\begin{equation}
\tilde{\rho}_{\beta}(u)=\lim_{\epsilon\to 0^+}\frac{1}{\beta N}\frac{\partial}{\partial t}\ln Z^{(\beta)}_\epsilon(t,J;u)\Big|_{t=0}\ .
\end{equation}

Averaging now over the matrix ensemble
\begin{equation}
\rho_{\beta}(u)=\left\langle\tilde{\rho}_{\beta}\left(u\right)\right\rangle_{J}
\end{equation}
and sending $\beta\to\infty$ at the very end, the density of the top eigenvector's components is eventually given by the remarkable formula
\begin{equation}
\rho(u)=\lim_{\beta\to\infty}\lim_{\epsilon\to 0^+}\frac{1}{\beta N}\frac{\partial}{\partial t}\left\langle\ln Z^{(\beta)}_\epsilon(t,J;u)\right\rangle_J\Big|_{t=0}\ .\label{remarkable}
\end{equation}

To compute the average of the logarithm of the auxiliary partition function
$Z^{(\beta)}_\epsilon(t,J;u)$, we will employ the replica trick once again
\begin{equation}
\left\langle\ln Z^{(\beta)}_\epsilon(t,J;u)\right\rangle_J=\lim_{n\to 0}\frac{1}{n}\ln \left\langle [Z^{(\beta)}_\epsilon(t,J;u)]^n\right\rangle_J\ .\label{replicavecZeps}
\end{equation}

We can anticipate that the replicated partition function will take the form

\begin{equation}
 \left\langle [Z^{(\beta)}_\epsilon(t,J;u)]^n\right\rangle_J\propto \int\mathcal{D}\varphi\mathcal{D}\hat{\varphi}\mathrm{d}\vec{\lambda}\exp\left [NS^{(\beta)}_{n}\left[\varphi,\hat{\varphi},\vec{\lambda};t,\epsilon;u\right]\right]\ ,
\end{equation}
where $\varphi$ and $\hat\varphi$ are functional order parameters. In a saddle point approximation for large $N$
\begin{equation}
 \left\langle [Z^{(\beta)}_\epsilon(t,J;u)]^n\right\rangle_J\approx \exp\left [NS^{(\beta)}_{n}\left(\varphi^\star,\hat{\varphi}^\star,\vec{\lambda}^\star;t,\epsilon;u\right)\right]\ ,\label{actionvec}
\end{equation}
where the starred objects satisfy self-consistency equations in which $t$ can be safely set to zero: indeed, the partial derivative $\frac{\partial}{\partial t}$ in \eqref{remarkable} only acts on terms containing any \emph{explicit} dependence on $t$, and not through any other indirect functional dependence. Inserting \eqref{actionvec} into \eqref{replicavecZeps}, and assuming that
\begin{equation}
S^{(\beta)}_{n}\left(\varphi^\star,\hat{\varphi}^\star,\vec{\lambda}^\star;t,\epsilon;u\right)\sim n s_\beta\left(t,\epsilon;u\right) +o(n)\label{sbeta}
\end{equation}
as $n\to 0$ (in a replica-symmetric setting), the final expression for the average density of top eigenvector's components for $N\to\infty$ reduces to
\begin{equation}
\rho(u)=\lim_{\beta\to\infty}\frac{1}{\beta}s_\beta'\left(0,0;u\right)\ ,\label{betalimitfinal}
\end{equation}
where $(\cdot)'$ stands for differentiation with respect to $t$.

Interestingly, we will find that the stationarity conditions defining
$\varphi^\star$, $\hat{\varphi}^\star$ and $\lambda^\star$ at the saddle point
for $t=0$ are just identical to those found in the replica calculation
for the largest eigenvalue. The explicit $n$-dependence of the action $S^{(\beta)}_{n}\left(\varphi^\star,\hat{\varphi}^\star,\vec{\lambda}^\star;t,\epsilon;u\right)$ is extracted by representing
the order parameters $\varphi$ and $\hat{\varphi}$ as an infinite superposition of
Gaussians, as previously done for the leading eigenvalue calculation. 

In the next subsections, we will apply this formalism to weighted Erd\H{o}s-R\'enyi  and random regular adjacency matrices.
 
\subsubsection{Erd\H{o}s-R\'enyi graph: weighted adjacency matrix.}

The average replicated partition function becomes

\begin{align}
\nonumber & \left\langle [Z^{(\beta)}_\epsilon(t,J;u)]^n\right\rangle_J  =\int\prod_{a=1}^{n}\left(\mathrm{d}\bm{v}_{a}\right)\int\prod_{a=1}^{n}\left(\frac{\beta}{4\pi}\mathrm{d}\lambda_{a}\right)\exp\left( \mathrm{i}\frac{\beta}{2}N\sum_{a}\lambda_{a}\right) \\
 & \times\exp\left[ \frac{c}{2N}\sum_{ij}\left(\left\langle \mathrm{e}^{\beta K\sum_{a}v_{ia}v_{ja}}\right\rangle_K-1\right)-\mathrm{i}\frac{\beta}{2}\sum_{a}\sum_{i}\lambda_{a}v_{ia}^{2}+\beta t\sum_{a}\sum_{i}\delta_\epsilon\left(u-v_{ia}\right)\right]\ ,
\end{align}
in complete analogy with \eqref{replicatedeig}. 

By introducing the functional order parameter

\begin{equation}
\varphi(\vec{v})=\frac{1}{N}\sum_{i}\prod_{a}\delta\left(v_{a}-v_{ia}\right)
\end{equation}

via the usual integral identity
\begin{equation}
1=\int N\mathcal{D}\varphi\mathcal{D}\hat{\varphi}\exp\left\{ -\mathrm{i}\int\mathrm{d}\vec{v}\hat{\varphi}(\vec{v})\left[N\varphi(\vec{v})-\sum_{i}\prod_{a}\delta\left(v_{a}-v_{ia}\right)\right]\right\}\ ,
\end{equation}
the replicated partition function can be once again cast in a form that
allows for a saddle point approximation
\begin{equation}
 \left\langle [Z^{(\beta)}_\epsilon(t,J;u)]^n\right\rangle_J\propto \int\mathcal{D}\varphi\mathcal{D}\hat{\varphi}\mathrm{d}\vec{\lambda}\exp\left [NS^{(\beta)}_{n}\left[\varphi,\hat{\varphi},\vec{\lambda};t,\epsilon;u\right]\right]\ ,
\end{equation}
where the action $S^{(\beta)}_{n}\left[\varphi,\hat{\varphi},\vec{\lambda};t,\epsilon;u\right]$ is the sum of four terms
\begin{equation}
S^{(\beta)}_{n}\left[\varphi,\hat{\varphi},\vec{\lambda};t,\epsilon;u\right] =S_{1}[\varphi,\hat{\varphi}]+S_{2}[\varphi]+S_{3}(\vec{\lambda})+S_{4}[\hat{\varphi},\vec{\lambda};t,\epsilon;u]\ ,
\end{equation}
where for simplicity we omit the full dependence on variables on the right hand side. The first three contributions are identical to those appearing in the largest eigenvalue calculation (see \eqref{eq:S1}, \eqref{eq:S2} and \eqref{eq:S3}). The explicit $t$ and $u$ dependence is confined to the  fourth contribution,

\begin{equation}
S_{4}[\hat{\varphi},\vec{\lambda};t,\epsilon;u] =  \mathrm{Log}\int\mathrm{d}\vec{v}\exp\left[ -\mathrm{i}\frac{\beta}{2}\sum_{a}\lambda_{a}v_{a}^{2}+\beta t\sum_{a}\delta_\epsilon\left(u-v_{a}\right)+\mathrm{i}\hat{\varphi}\left(\vec{v}\right)\right]\ .\label{S4mu}
\end{equation}

The saddle point equations for $\varphi^\star,\hat{\varphi}^\star$ (where we can safely set $t=0$)  are then identical to those (see \eqref{eq:rho_stat} and \eqref{eq:hat_rho_stat}) appearing in the calculation for the average largest eigenvalue. Therefore we can follow the same strategy as before, and represent $\varphi^\star$ and $\hat\varphi^\star$ as uncountably infinite superposition of Gaussians, whose parameters fluctuate according to joint pdfs $\pi$ and $\hat\pi$ as in \eqref{eq:ansatz_rep_1} and \eqref{eq:ansatz_rep_2}. Such joint pdfs satisfy the very same self-consistency equations as in \eqref{eq:pi} and \eqref{eq:pi_hat} and for these reasons we can use the same labels as before. The only difference with respect to the previous case is in the extra $t$-derivative that we have to take from the contribution $S_4(\hat\varphi^\star,\lambda;t,\epsilon;u)$.

Inserting the ansatz 
\begin{equation}
\mathrm{i}\hat{\varphi}^\star(\vec{v}) =\hat{c}\int\mathrm{d}\hat{\omega}\mathrm{d}\hat{h}\ \hat{\pi}(\hat{\omega},\hat{h})\prod_{a=1}^{n}\exp\left( \frac{\beta}{2}\hat{\omega}v_{a}^{2}+\beta\hat{h}v_{a}\right)
\end{equation}
into \eqref{S4mu}, and expanding $\mathrm{e}^{\mathrm{i}\hat{\varphi}^\star(\vec{v}) }=\sum_{s\geq 0}\frac{(\mathrm{i}\hat{\varphi}^\star(\vec{v}))^s}{s!}$, we obtain (in the limit $n\to 0$)
\begin{align}
\nonumber S_{4}(\hat{\varphi}^\star,\lambda^\star;t,\epsilon;u) &=  \hat{c}+n\sum_{s=0}^{\infty}p_{\hat{c}}\left(s\right)\int\{ \mathrm{d}\hat{\pi}\} _{s}~\mathrm{Log}\int\mathrm{d}v\exp\left[ -\mathrm{i}\frac{\beta}{2}\lambda^\star v^{2}+\beta t\delta_\epsilon\left(u-v\right)
\right.\\
&\left.+ \frac{\beta}{2}\{\hat{\omega}\}_s v^{2}+\beta\{\hat{h}\}_s v\right]\ .
\end{align}
Therefore, we can isolate the function $s_\beta(t,\epsilon;u)$ in \eqref{sbeta} as
\begin{align}
\nonumber s_\beta(t,\epsilon;u) &=  \sum_{s=0}^{\infty}p_{c}\left(s\right)\int\{ \mathrm{d}\hat{\pi}\} _{s}~\mathrm{Log}\int\mathrm{d}v\exp\left[ -\frac{\beta}{2}\lambda v^{2}+\beta t\delta_\epsilon\left(u-v\right)
\right.\\
&\left.+ \frac{\beta}{2}\{\hat{\omega}\}_s v^{2}+\beta\{\hat{h}\}_s v\right]\ ,
\end{align}
in view of the identifications $\hat{c}=c$ and $\mathrm{i}\lambda^\star\equiv\lambda$ as before. Taking the $t$-derivative and setting $t$ and $\epsilon$ to zero, we get
\begin{align}
\nonumber s^\prime_\beta(0,0;u) &= \beta\sum_{s=0}^{\infty}p_{c}\left(s\right)\int\{ \mathrm{d}\hat{\pi}\} _{s}\frac{\exp\left[ -\frac{\beta}{2}(\lambda-\{\hat{\omega}\}_s) u^{2}+\beta\{\hat{h}\}_s u\right]}{\int\mathrm{d}v\exp\left[ -\frac{\beta}{2}(\lambda-\{\hat{\omega}\}_s)v^{2}+\beta\{\hat{h}\}_s v\right]}\ .\end{align}

Taking the $\beta\to\infty$ limit as in \eqref{betalimitfinal}, we eventually find

\begin{equation}
\rho(u)  =\sum_{s=0}^{\infty}p_{c}(s)\int\left\{\mathrm{d}\hat{\pi}\right\} _{s}\delta\left(u-\frac{\{ \hat{h}\} _{s}}{\lambda-\{ \hat{\omega}\} _{s}}\right)\ .\label{eq:density_vector_general}
\end{equation}
Expressing everything in terms of the $\pi$-distribution, indicating with $p_c(s)$ the actual degree distribution \eqref{eq:degree_pmf} and truncating the series at the largest degree $k_{\mathrm{max}}$ (as we did in previous sections), we eventually obtain

\begin{equation}
\rho(u)=\sum_{s=0}^{k_{\mathrm{max}}}p_{c}(s)\int\left\{\mathrm{d}\pi\right\} _{s}\left\langle\delta\left(u-\frac{\sum_{\ell=1}^{s}\frac{h_{\ell}K_{\ell}}{\omega_{\ell}}}{\lambda-\sum_{\ell=1}^{s}\frac{K_{\ell}^2}{\omega_{\ell}}}\right)\right\rangle_{\{K\}_s}\ ,\label{eq:density_with_pi}
\end{equation}
where $\pi(\omega,h)$ satisfies the self-consistent equation \eqref{eq:pi_trunc} (to be solved via population dynamics), supplemented with the normalisation condition \eqref{eq:lambdastarconditionV2}. Once again, the brackets $\left\langle\cdot \right\rangle _{\{ K\}_s }$ denote averaging w.r.t to a collection of $s$ i.i.d random variables $K_{\ell}$, each drawn from the bond weight pdf $p(K)$.

Eq. \eqref{eq:density_with_pi} essentially recovers Eq. \eqref{eq:first_eigv_comp_denisity_cavity} found with the cavity
method. As a general remark, it is worth noticing that the $\beta$-dependent distribution $\rho_{\beta}(u)$ had already arisen naturally in the eigenvalue calculation when evaluating the stationarity conditions
with respect to $\lambda$. In fact, the distribution in \eqref{eq:p_bar} is exactly identical to $\rho_{\beta}(u)$. Moreover, in the cavity formalism, $\rho_{\beta}\left(u\right)$ is closely related to the single-site marginal of a single instance \eqref{eq:single_site_marginals}.

We remark once again that -- in analogy with the typical largest eigenvalue calculation -- the validity of Eq. \eqref{eq:density_with_pi} is not restricted to a truncated Poisson degree distribution \eqref{eq:degree_pmf}. It actually provides the density of the top eigenvector's components for the weighted adjacency matrix of \emph{any} 
configuration model with finite connectivity and bounded maximal degree as a weighted superposition of delta functions, one for each degree of the graph. It is then natural to identify the quantity $\frac{\sum_{\ell=1}^{s}\frac{h_{\ell}K_{\ell}}{\omega_{\ell}}}{\lambda-\sum_{\ell=1}^{s}\frac{K_{\ell}^2}{\omega_{\ell}}}$ as the contribution to the density coming from nodes of degree $s$.

\begin{figure}[!b]
\begin{centering}
\includegraphics[scale=0.5]{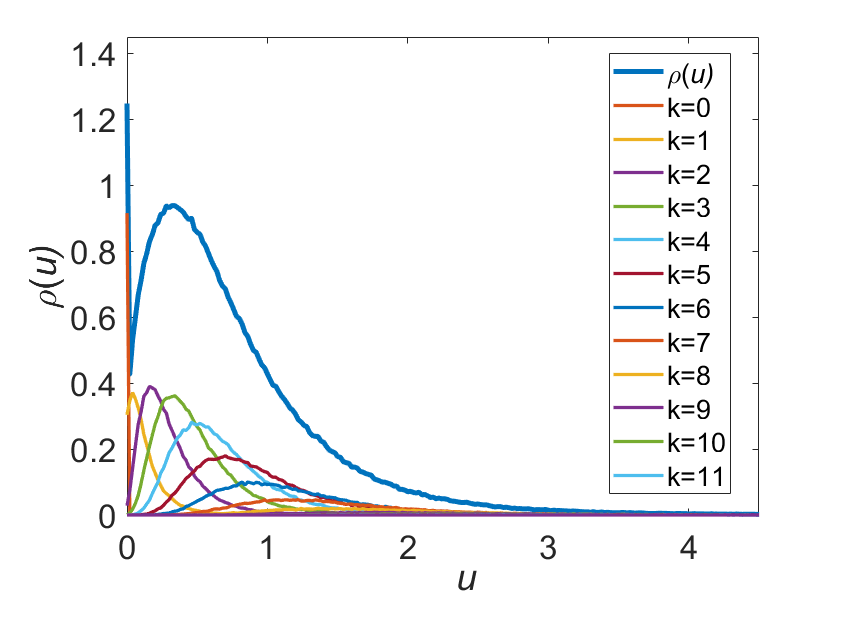}
\includegraphics[scale=0.5]{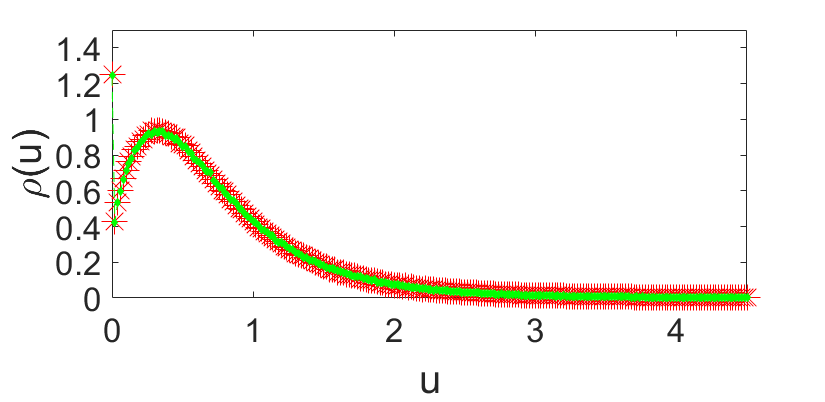}
\par\end{centering}

\begin{centering}
\caption{Density of the top eigenvector components $\rho(u)$  for Erd\H{o}s-R\'enyi weighted adjacency matrices in the limit $N\rightarrow\infty$. The graphs are obtained via the population dynamics algorithm. As in Fig. \ref{fig:weightedER}, the mean connectivity is $c=4$, $k_{\mathrm{max}}=16$ and the population size is $N_{P}=10^{6}$. The bond weight distribution is chosen to be uniform, specifically $p(K)=1/2$ for all $K\in[1,3]$.  {\bf Top panel}: the thick blue line represents  the full distribution $\rho(u)$, whereas the thinner curves underneath indicate the various degree contributions $k=0,1,2,3,..$. Once again, the peak at $u=0$ is given by the contribution of isolated nodes ($k=0$). Larger degree nodes contribute to the tail of the distribution. Once again, only the degree contributions uo to $k=11$ has been labelled. {\bf Bottom panel}: the comparison between results for the density of components \eqref{eq:first_eigv_comp_denisity_cavity}
or equivalently \eqref{eq:density_with_pi} of the top eigenvector
obtained with population dynamics (red stars) and results obtained
with direct diagonalisation (green diamonds) shows perfect agreement between the two.  }\label{fig:weighted_density}
\par\end{centering}
\centering{}
\end{figure}

The $s=0$ contribution from isolated nodes indeed gives rise to the sharp peak at $u=0$.
The $\rho(u)$ of a Erd\H{o}s-R\'enyi $\{0,1\}$-adjacency matrix is shown in Figure \ref{fig:summary} (top panels), whereas the case of weighted Erd\H{o}s-R\'enyi  adjacency matrices is shown in Figure \ref{fig:weighted_density}.

\subsubsection{Random regular graph: adjacency matrix}\label{sec:rrg2}
In this case, building on subsection \ref{sec:rrg} and recalling that $p_c(s)=\delta_{s,c}$ and $p(K)=\delta(K-1)$,
the ratio in \eqref{eq:density_with_pi} simply becomes $c(c-2)/[c(c-1)-c]=1$, entailing
\begin{equation}
\rho(u)=\delta\left(u-1\right),
\end{equation}
as expected.

\subsubsection{Large-$c$ limit for weighted adjacency matrices}\label{sec:dense}
We consider now the large $c$-limit of Erd\H{os}-R\'enyi graphs (more generally, any configuration model graph for which  $\frac{\sigma_k^2}{\langle k\rangle^2} = \frac{\langle k^2\rangle-\langle k\rangle^2}{\langle k\rangle^2} \to 0$ as  $\langle k\rangle = c \to \infty$). A meaningful large-$c$ limit is obtained for Eq. \eqref{eq:pi} by rescaling each instance of the bond random weights as $K_{ij} = \mathcal{J}_{ij}/\sqrt{c}$, leading to
\begin{equation}
\pi(\omega,h)=\sum_{s\geq 1}\frac{s}{c}p_{c}(s)\int\{\mathrm{d}\pi\} _{s-1}\left\langle \delta\left(\omega-\lambda+\frac{1}{c}\sum_{\ell=1}^{s-1}\frac{\mathcal{J}_{\ell}^{2}}{\omega_{\ell}}\right)\delta\left(h-\frac{1}{\sqrt{c}}\sum_{\ell=1}^{s-1}\frac{h_{\ell}\mathcal{J}_{\ell}}{\omega_{\ell}}\right)\right\rangle _{\{ \mathcal{J}\}_{s-1} }\ .\label{eq:pi_rescaled}
\end{equation}
In the $c\gg 1$ limit, the $s$-sum in Eq. \eqref{eq:pi_rescaled} is 
restricted to $s = c  \pm \mathcal{O}(\sigma_k)$ (with $\sigma_k = \sqrt{c}$ for Erd\H{os}-R\'enyi  graphs), so that the argument appearing
in the first $\delta$-function on the r.h.s of this equation can be evaluated by appeal to the Law of Large Number (LLN). This entails that
\begin{equation}
\omega = \lambda - \frac{1}{c}\sum_{\ell=1}^{s-1}\frac{\mathcal{J}_{\ell}^{2}}{\omega_{\ell}}
\end{equation}
is {\em non-fluctuating\/}, hence the self-consistency equation demands that 
\begin{equation}
\pi\left(\omega,h\right)= \delta(\omega -\bar\omega) \times P(h)\ ,
\end{equation}
with (by the LLN)
\begin{equation}
 \bar{\omega} = \lambda - \frac{1}{c}\sum_{\ell=1}^{s-1}\frac{\mathcal{J}_{\ell}^{2}}{\bar\omega} = \lambda - \frac{\langle \mathcal{J}^2\rangle_{\mathcal{J}}}{\bar{\omega}}\ .
\end{equation}
Specializing to $\langle \mathcal{J}^2\rangle_{\mathcal{J}} = 1$, we see that 
\begin{equation}
\bar\omega_{1,2} = \frac{1}{2}\left(\lambda \pm \sqrt{\lambda^2 - 4}\right)\ ,
\end{equation}
which requires $\lambda \ge 2$ to have real positive $\bar\omega$.

Similarly, the argument of the second $\delta$-function on the r.h.s of \eqref{eq:pi_rescaled} exhibits a scaling that allows one to 
conclude (for $\langle \mathcal{J}_\ell\rangle_{\mathcal{J}} =0$) that
$$
h = \frac{1}{\sqrt{c}}\sum_{\ell=1}^{k-1}\frac{h_{\ell}\mathcal{J}_{\ell}}{\omega_{\ell}}=
\frac{1}{\sqrt{c}}\sum_{\ell=1}^{k-1}\frac{h_{\ell}\mathcal{J}_{\ell}}{\bar \omega}
 \sim {\cal N}(0,\sigma_h^2)
$$
by appeal to the Central Limit Theorem. The variance follows using independence of the $\{h_\ell\}$ and $\{\mathcal{J}_\ell\}$
\begin{equation}
\sigma_h^2 = \langle h^2 \rangle = \frac{1}{c \bar\omega^2}\sum_{\ell=1}^{s-1} \langle h_\ell^2\rangle \langle \mathcal{J}_\ell^2\rangle_{\mathcal{J}} = \frac{\sigma_h^2}{\bar\omega^2} \ .\label{sigmahsquare}
\end{equation}
This equation allows a finite variance if and only if $\bar\omega^2 =1$, which requires $\lambda =\pm 2$, i.e. that $\lambda$ -- the most probable location of the largest eigenvalue --
is at the edge of the Wigner semi-circle (and we require the positive solution).

To obtain the distribution $\rho(u)$ of eigenvector components, it is instructive and more direct to look back at the cavity equations \eqref{eq:Q}, \eqref{eq:first_eigv_comp_denisity_cavity} and \eqref{eq:cavity_normalization}. After the rescaling $K_\ell = \mathcal{J}_\ell/\sqrt{c}$ and in the large $c$-limit, it is easy to see from \eqref{eq:Q} that $\Omega =\bar\omega$ and that $H$ is a sum of Gaussians, and thus itself Gaussian, of
variance $\sigma_h^2/\bar\omega^2\equiv \sigma_h^2$ by \eqref{sigmahsquare}. It then follows from the normalisation condition \eqref{eq:cavity_normalization} that $\sigma_h^2 = 1$, so eventually 
\begin{equation}
\rho(u) = \frac{1}{\sqrt{2\pi}} {\rm e}^{-u^2/2}\ .
\end{equation}
Looking now at the variable $\eta=u^2$, and noting that positive and negative $u$
give rise to the same $\eta$, one obtains by the simple transformation of pdf's
\begin{equation}
\rho(\eta) = \frac{1}{\sqrt{2\pi \eta}} {\rm e}^{-\eta/2}\ ,
\end{equation}
which is the standard form of Porter-Thomas distribution for real-valued (invariant) random matrices (see \cite{Livan2018}, Eq. (9.10)).

\section{Application: sparse random Markov transition matrices \label{sec:Markov}}
In this section, we cross-check the formalism with an ensemble of transition matrices $W$ for discrete Markov chains in an $N$-dimensional state space. The evolution equation for the probability vector $\bm{p}(t)$ is given by
\begin{equation}
\bm{p}(t+1)=W\bm{p}(t)\ .
\end{equation}
The transition matrix $W$ is such that $W_{ij}\geq0\ \forall(i,j)$ and $\sum_{i}W_{ij}=1\ \forall j$. For an irreducible chain, the top right eigenvector of the matrix $W$  corresponding to the Perron-Frobenius eigenvalue $\lambda_1=1$ represents the unique equilibrium distribution, i.e. $\bm{v}_1=\bm{p}^{\mathrm{eq}}$.
The matrix $W$ is in general not symmetric: however, if the Markov process satisfies a detailed balance condition, i.e. $W_{ij}p_j^{\mathrm{eq}}=W_{ji}p_i^{\mathrm{eq}}$, it can be symmetrised via a similarity transformation, yielding
\begin{equation}
W_{ij}^S=(p_i^{\mathrm{eq}})^{-1/2}W_{ij}(p_j^{\mathrm{eq}})^{1/2}\ .
\end{equation}

The symmetrised matrix $W^S$ will be the target of our analysis: even though it is not itself a Markov matrix since the columns normalisation constraint is lost, in view of the detailed balance condition $W^S$ has the same (real) spectrum of $W$, and its top eigenvector $\bm{v}_1$ is given in terms of the top right eigenvector of $W$, $\bm{p}_{\mathrm{eq}}$, as
\begin{equation}
v_1^{(i)}=(p_i^{\mathrm{eq}})^{1/2}\ .
\end{equation}

We will consider  the case of an unbiased random walk: the matrix $W$ is then defined as
\begin{equation}
W_{ij}=
\begin{cases}
\frac{c_{ij}}{k_j}, & i\neq j \\
1, & i=j\ \mathrm{and}\ k_j=0\ ,
\end{cases}
\end{equation}
where $c_{ij}$ represents the connectivity matrix and $k_j=\sum_i c_{ij}$ is the degree of the node $j$. In this case, the top right eigenvector of $W$ is proportional to the vector expressing the degree sequence: for our purposes, we choose  the inverse of the mean degree as proportionality constant, i.e. $p_i^{\mathrm{eq}}=k_i/\left\langle k \right\rangle$. The symmetrised matrix $W^S$ is expressed as
\[
W_{ij}^S=
\begin{cases}
\frac{c_{ij}}{\sqrt{k_i k_j}}, & i\neq j \\
1, & i=j\ \mathrm{and}\ k_j=0\ ,
\end{cases}
\]
with its top eigenvector being $v_1^{(i)}=\sqrt{k_i/\left\langle k \right\rangle}$. Therefore, we expect that  
\begin{equation}
\rho(u)=\sum_{k\geq k_{\mathrm{min}}} p(k) \delta\left( u-\sqrt{\frac{k}{\left\langle k \right\rangle}} \right)\ , \label{eq:density_markov_theory}
\end{equation}
where $p(k)$ is the degree distribution of the connectivity matrix $\{c_{ij}\}$.

In order to avoid isolated nodes and isolated clusters of nodes, we consider a shifted Poissonian degree distribution with $k_{\mathrm{min}}=2$, i.e. 
\begin{equation}
p(k)=\frac{\mathrm{e}^{-c}c^{k-2}}{\left( k-2\right)!}\mathbbm{1}_{k\geq2}\ ,
\end{equation}
with mean degree $\left\langle k \right\rangle=c+2$. 

The single-instance cavity treatment starts from the Gibbs-Boltzmann distribution 
\begin{equation}
P_{\beta,W^S}\left(\bm{v}\right)=\frac{1}{Z}\exp\left\{ \beta\left[\frac{1}{2}\sum_{ij}^N v_i \frac{c_{ij}}{\sqrt{k_i k_j}}v_j-\frac{\lambda}{2}\sum_{i}^N v_i^2\right]\right\} \ ,
\end{equation}
which, after the change of variable $\tilde{v_i}=v_i/\sqrt{k_i}$, becomes
\begin{equation}
P_{\beta,W^S}\left(\tilde{\bm{v}}\right)=\frac{1}{Z}\exp\left\{ \beta\left[\frac{1}{2}\sum_{ij}^N \tilde{v_i} c_{ij}\tilde{v_j}-\frac{\lambda}{2}\sum_{i}^N k_i {\tilde{v_i}}^2\right]\right\} \ .
\end{equation}

It is convenient to frame and solve the problem in terms of the vector $\tilde{\bm{v}}$, since in this case the matrix involved in the analysis is just the standard $\{0,1\}$-adjacency matrix of the underlying graph, as in \cite{Kuehn2015,Kuehn2015a}.
The cavity single-instance equations for this problem read 
\begin{align}
\Omega_{j}^{(i)} &=\lambda k_j-\sum_{\ell\in\partial j\backslash i}\frac{1}{\Omega_{\ell}^{(j)}}\ ,\label{eq:A_markov}\\
H_{j}^{(i)} &=\sum_{\ell\in\partial j\backslash i}\frac{H_{\ell}^{(j)}}{\Omega_{\ell}^{(j)}}\ ,\label{eq:H_markov}
\end{align}
whereas the equations for the single-site marginal coefficients read 
\begin{align}
\Omega_{i} &=\lambda k_i-\sum_{j\in\partial i}\frac{1}{\Omega_{j}^{(i)}}\,,\label{eq:A_marg_markov}\\
H_{i} &=\sum_{j\in\partial i}\frac{H_{j}^{(i)}}{\Omega_{j}^{(i)}}\,.\label{eq:H_marg_markov}
\end{align}

In the thermodynamic limit $N\to\infty$, the equations \eqref{eq:A_markov} and \eqref{eq:H_markov} lead to
\begin{equation}
q\left(\omega,h\right)=\sum_{k=2}^{\infty}\frac{k}{\left\langle k \right\rangle}p\left(k\right) \int\left[\prod_{\ell=1}^{k-1}\mathrm{d}q\left(\omega_{\ell},h_{\ell}\right)\right] \delta\left(\omega-\lambda k+\sum_{\ell=1}^{k-1}\frac{1}{\omega_{\ell}}\right)\delta\left(h-\sum_{\ell=1}^{k-1}\frac{h_{\ell}}{\omega_{\ell}}\right)\ ,\label{eq:q_markov}
\end{equation}
in complete analogy with \eqref{eq:q}.

Similarly,  equations \eqref{eq:A_marg_markov} and \eqref{eq:H_marg_markov} lead to
\begin{equation}
Q\left(\Omega,H\right)=\sum_{k=2}^{\infty}p\left(k\right)\int\left[\prod_{\ell=1}^{k}\mathrm{d}q\left(\omega_{\ell},h_{\ell}\right)\right] \delta\left(\Omega-\lambda k+\sum_{\ell=1}^{k}\frac{1}{\omega_{\ell}}\right)\delta\left(H-\sum_{\ell=1}^{k}\frac{h_{\ell}}{\omega_{\ell}}\right)\ ,\label{eq:Q_markov}
\end{equation}
entailing that the density of the top eigenvector's components in the space of vectors $\tilde{\bm{v}}$  is given by
\begin{equation}
\rho\left(\tilde{u}\right)=\int\mathrm{d}\Omega\mathrm{d}HQ\left(\Omega,H\right)\delta\left(\tilde{u}-\frac{H}{\Omega}\right)\ ,\label{eq:first_eigv_markov}
\end{equation} 
which follows from the general theory. 

\begin{figure}[!hb]
\begin{centering}
\includegraphics[scale=0.5]{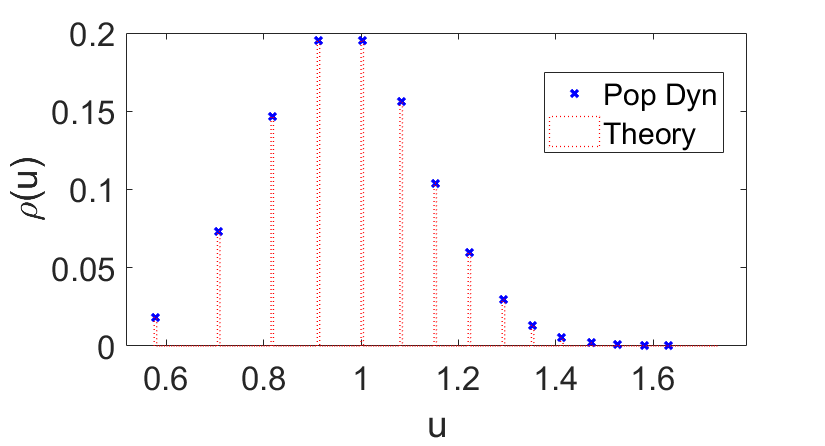}

\par\end{centering}

\begin{centering}
\caption{Density of the top eigenvector's components for sparse Markov matrices representing the transition matrices of unbiased random walks in the thermodynamic limit $N\to\infty$.  The histogram has been produced by population dynamics with a population of size $N_P=10^6$, specialised to the case of a shifted Poissonian degree distribution with minimum degree $k_{\mathrm{min}}=2$ and average degree $\left\langle k \right\rangle = 6$ ($c=4$). 
The simulation results (blue crosses) match the theoretical predictions (red dashed bars).}
\label{fig:markov}
\par\end{centering}
\centering{}
\end{figure}

As before, \eqref{eq:q_markov} and \eqref{eq:Q_markov} are efficiently solved via a population dynamics algorithm: as expected, the convergence is attained for $\lambda=1$, i.e. in correspondence of the largest eigenvalue of $W^S$. 
Running the simulations, we find that the distribution $\rho(\tilde{u})$ converges to a delta peak centered at a finite real positive value: this behaviour agrees perfectly with the theoretical predictions, because it precisely implies that $\rho(u)$ must be given by \eqref{eq:density_markov_theory}. Indeed, the two quantities are related via the aforementioned change of variables, $u\leftarrow\tilde{u}\sqrt{k}$, and the constant value the variables $\tilde{u}$ converge to corresponds to $1/\sqrt{\left\langle k \right\rangle}$, once the normalisation is fixed according to \eqref{eq:cavity_normalization}. In Fig. \ref{fig:markov}, we compare the density of the top eigenvector's components for sparse Markov matrices (representing the transition matrices of unbiased random walks) with numerical diagonalisation.

As a concluding remark, we notice that the same route can be followed to characterise the top eigenpair statistics of the so-called \emph{tilted} Markov transition matrix \cite{Touchette2009} appearing in the context of rare events for random walks on networks \cite{DeBacco2016}. This will be discussed in a separate publication.

\section{Population dynamics\label{sec:Population-Dynamics}}

The population dynamics algorithm employed to solve \eqref{eq:pi_trunc}
is deeply rooted in the statistical mechanics of spin glasses \cite{Mezard2001,Krzakala2016}.
In our context, it can be summarised as follows. 

Two coupled populations with $N_{P}$ members each $\left\{ \left(\omega_{i},h_{i}\right)\right\} _{1\leq i\leq N_{P}}$
are randomly initialised, taking into account that $\omega_{i}>\zeta$, where $\zeta$ is the upper edge of the support of the pdf $p(K)$.

For any suitable value of $\mathrm{i}\lambda^*\equiv\lambda\in\mathbb{R}$, the following steps are iterated until stable populations are obtained:

\begin{enumerate}
\item Generate a random $s\sim\frac{s}{c}p_{c}\left(s\right)$, where $c=\left\langle s \right\rangle$
\item Generate $s-1$ i.i.d random variables $K_{\ell}$ from the bond weights pdf $p(K)$
\item Select $s-1$ pairs $\left(\omega_{\ell},h_{\ell}\right)$ from
the population at random; compute
\begin{align}
\omega^{(new)} & =  \lambda-\sum_{\ell=1}^{s-1}\frac{K_{\ell}^2}{\omega_{\ell}}\ ,\\
h^{(new)} & =  \sum_{\ell=1}^{s-1}\frac{h_{\ell}K_{\ell}}{\omega_{\ell}}\ ,
\end{align}
and replace a randomly selected pair $\left(\omega_{j},h_{j}\right)$
where $j=1,...,N_{P}$ with the pair $\left(\omega^{(new)},h^{(new)}\right)$. 
\item Return to (i).
\end{enumerate}
Convergence is assessed by looking at the first moments of the vector formed by the $N_P$ samples. A \emph{sweep} is completed when all the $N_P$ pairs $(\omega_j,h_j)$ of the population have been updated at least once according to the steps above.

The procedure to evaluate \eqref{eq:Q} (or alternatively \eqref{eq:first_eigv_comp_denisity_cavity}) is almost identical, except for the details concerning the $s$-sampling. Starting from two coupled populations with $N_P$ members $\{(\Omega_i,H_i)\}_{1\leq i\leq N_P}$, the following steps are iterated:
\begin{enumerate}
\item Generate a random $s\sim p_{c}\left(s\right)$, where $c=\left\langle s \right\rangle$
\item Generate $s$ i.i.d random variables $K_{\ell}$ from the bond weights pdf $p(K)$
\item Select $s$ pairs $\left(\omega_{\ell},h_{\ell}\right)$ from
the population $\left\{ \left(\omega_{i},h_{i}\right)\right\} _{1\leq i\leq N_{P}}$ at random; compute
\begin{align}
\Omega^{(new)} & =  \lambda-\sum_{\ell=1}^{s}\frac{K_{\ell}^2}{\omega_{\ell}}\ ,\\
H^{(new)} & =  \sum_{\ell=1}^{s}\frac{h_{\ell}K_{\ell}}{\omega_{\ell}}\ .
\end{align}
\item Replace a randomly selected pair $\left(\Omega_{j},H_{j}\right)$
where $j=1,...,N_{P}$ with the pair $\left(\Omega^{(new)},H^{(new)}\right)$, which is then a new sample from $Q(\Omega,H)$. It can be used via Eq. \eqref{eq:first_eigv_comp_denisity_cavity} to create $u^{(new)}=H^{(new)}/\Omega^{(new)}$ as a new sample from $\rho(u)$.
\item Return to (i).
\end{enumerate}

The value of the parameter $\lambda$ controls the convergence of the algorithm: indeed, the convergence to a non-trivial distribution is achieved only when $\lambda$ is equal to the typical largest eigenvalue $\left\langle\lambda_1 \right\rangle_J$, as prescribed by the theory: for any $\lambda>\left\langle\lambda_1 \right\rangle_J$, the variables of type $h$ will shrink to zero, whereas for $\lambda<\left\langle\lambda_1 \right\rangle_J$ they will blow up in norm. Hence, the value  $\lambda=\left\langle\lambda_1 \right\rangle_J$ is the only value for which the normalisation condition \eqref{eq:lambdastarconditionV2} (or equivalently \eqref{eq:cavity_normalization}) can be satisfied, in complete agreement with the replica predictions.

In view of the expected behaviour described above, we will initially start from a large value of $\lambda$, which is then progressively decreased until convergence is achieved. A suitable starting value is given by the largest degree $k_\mathrm{max}$ that appears in the connectivity distribution.
The value of $k_\mathrm{max}$ is fixed in such a way that $p_c(k_\mathrm{max})N_P\geq1$: only if this condition is met, the value $k_\mathrm{max}$ appears at least once in the degree array that is created to sample from $p_c(k)$. Because of this choice, the largest degree depends on the limits of the machine that is used to run the population dynamics algorithm: by using a population size $N_P=10^6$ and a parameter $\bar{c}=4$ in \eqref{eq:degree_pmf}, we are able to reach $k_\mathrm{max}=16$. Thus, the normalisation constant $\mathcal{N}$ in \eqref{eq:degree_pmf} is not very different from $1$ and $\bar{c}\simeq c=\langle s\rangle$, making the truncation of the Poisson distribution - for all practical purposes - ineffective.


Once $\lambda$ has been set to the only value ($=\left\langle\lambda_1 \right\rangle_J$) for which a non-trivial finite normalisation can be found, the value of such normalisation can be adjusted by properly rescaling the $h$'s. Such rescaling is always allowed due to the linear nature of the recursion that governs their update. This recursion will be discussed in detail in \ref{sec:Non_backtracking}.

The population dynamics algorithm can also be employed to evaluate numerically the integral in \eqref{eq:ilambdastar}. The integral has the following structure:
\begin{equation}
I=\int\mathrm{d}\pi(\omega,h)\mathrm{d}\pi(\omega',h')\left\langle f\left(\omega,h,\omega',h',K\right)\right\rangle_{K}\ ,\label{eq:integral}
\end{equation}
where $f$ is a generic function of the cavity fields and $K$.
Once the correct value of $\lambda=\left\langle\lambda_1\right\rangle_J$ has been found, a number $N_{\mathrm{eq}}$ of equilibration sweeps is performed, following the protocol illustrated above.

 After equilibration, a variable $F=0$ is initialised. Then for $j=1,\ldots,N_{\mathrm{meas}}$:
\begin{enumerate}
\item Perform a sweep
\item Pick $(\omega,h)$ and $(\omega',h')$ at random, generate $K\sim p(K)$ and compute $f\left(\omega,h,\omega',h',K\right)$.
\item Update $F$: $F=F+f\left(\omega,h,\omega',h',K\right)$.
\end{enumerate}
The value of the integral \eqref{eq:integral} is approximated by invoking the law of large numbers, as
\begin{equation}
I\simeq\frac{F}{N_{\mathrm{meas}}}\ ,
\end{equation}
where the typical fluctuation is of the order of $1/\sqrt{N_P N_{\mathrm{meas}}}$.

\section{Conclusions}\label{sec:conclusions}

In summary, we have further developed a formalism - pioneered by Kabashima and collaborators -  to compute the statistics of the largest eigenvalue and of the corresponding top eigenvector for some ensembles of sparse symmetric matrices, i.e. (weighted) adjacency matrices of graphs with finite mean connectivity. The top eigenpair problem can be recast as the optimisation of a quadratic Hamiltonian on the sphere: introducing the associated Gibbs-Boltzmann distribution and a fictitious inverse temperature $\beta$, the top eigenvector represents  the ground state of the system, which is attained in the limit $\beta\to\infty$. In order to extract this limit, we have employed two methods, cavity and replicas, both borrowed from the Statistical Mechanics approach to disordered systems. 
We first analysed the case of a single-instance matrix within a ``grand canonical" cavity framework. The single-instance cavity method leads fairly quickly to superficially appealing recursion equations, however it has the obvious drawback of enlarging - and not reducing - the complexity of the problem: indeed, it turns a $N$-dimensional problem involving the single matrix $J$ into an $Nc$ dimensional problem - where
$c=\left\langle k\right\rangle>1$ is the mean degree - involving the non-backtracking operator $B$, as detailed in \ref{sec:Non_backtracking}. 

However, the cavity single-instance recursions constitute an essential ingredient to arrive at the equations \eqref{eq:q}, \eqref{eq:first_eigv_comp_denisity_cavity} and \eqref{eq:cavity_normalization} for the associated joint probability densities of the auxiliary fields of type $\Omega$ and $H$ that characterise the typical largest eigenvalue and the statistic of the top eigenvector in the thermodynamic limit $N\to\infty$. Moreover, the exact same equations (see \eqref{eq:pi_trunc}, \eqref{eq:density_with_pi} and \eqref{eq:lambdastarconditionV2}) are found via the completely alternative replica derivation, entailing that the two methods are equivalent in the thermodynamic limit.
Within the population dynamics algorithm employed to solve the stochastic recursion \eqref{eq:q} (or equivalently \eqref{eq:pi_trunc}), we are able to identify the typical largest eigenvalue as the parameter controlling the convergence of the algorithm, and unpack the contributions coming to nodes of different degrees to the average density of the top eigenvector's components. The simulations show excellent agreement of the theory with the direct diagonalisation of large matrices.
As a further cross-check of the formalism, we computed the average density of the top eigenvector's components of sparse Markov matrices representing unbiased random walks on a sparse network under the detailed balance condition, thus retrieving the expected relation between such components and the node degrees of the underlying network.

\ack{}{}

The authors acknowledge funding by the Engineering and Physical Sciences Research
Council (EPSRC) through the Centre for Doctoral Training in Cross Disciplinary Approaches to Non-Equilibrium Systems (CANES, Grant Nr. EP/L015854/1).

\appendix
\section{\\The solution of the single instance self-consistency equations and the non-backtracking operator.\label{sec:Non_backtracking}}

The set of self-consistency equations \eqref{eq:A} and \eqref{eq:H}
for the cavity fields, supplemented with \eqref{eq:A_marg} and \eqref{eq:H_marg} for the coefficients of the marginal distributions, constitutes the full solution of the top eigenpair problem for a single instance of a sparse matrix. 
Even in this case, the convergence of the update equations \eqref{eq:A} and \eqref{eq:H} is dictated by the value of the parameter $\lambda$, which once again is related to the possibility to normalise the resulting top eigenvector.

Note that \eqref{eq:H} is a linear recursion driven by the operator $B$, whose elements can be defined as
\begin{equation}
B_{(i,j),(k,\ell)}=\begin{cases}
\frac{J_{j\ell}}{\Omega_{\ell}^{(j)}} & i\neq\ell\land j=k\\
0 & \mathrm{otherwise}
\end{cases}\,.
\label{eq:non-back}
\end{equation}
$B$ is an example of non-backtracking operator, first introduced by Hashimoto in \cite{Hashimoto1989}. For a given graph, the Hashimoto non-backtracking operator $\tilde{B}$ in its original form counts the number of paths from a node $i$ to a node $\ell$ passing through a third node $j$, for every choice of these three different nodes. It is defined as

\begin{equation}
\tilde{B}_{(i,j),(k,\ell)}=\begin{cases}
1 & i\neq\ell\land j=k\\
0 & \mathrm{otherwise}
\end{cases}\ .
\end{equation}

In our case, if the absolute value of the largest eigenvalue of the modified non-backtracking operator $B$  is greater than $1$, the absolute values of the cavity fields $H_{j}^{(i)}$'s will blow up, whereas if it smaller than $1$, they will shrink to zero. Therefore, $\lambda$ must be tuned appropriately in \eqref{eq:A}
to prevent the linear recursion \eqref{eq:H} from landing on a
trivial solution. Indeed, when $\lambda$ is ``too large", the $\Omega_{j}^{(i)}$'s will be large too, resulting in a largest eigenvalue of  $B$  with magnitude smaller than $1$. This would suggest to progressively decrease $\lambda$ from a large value down to its lower bound $\lambda_{1}$, necessary to ensure that the optimisation problem is well-defined. In other words, the largest eigenvalue of the operator $B$ must be exactly $1$ for the $H_{j}^{(i)}$'s to have a finite norm. This will happen only when $\lambda=\lambda_{1}$.

\begin{figure}
\begin{centering}
\includegraphics[scale=1]{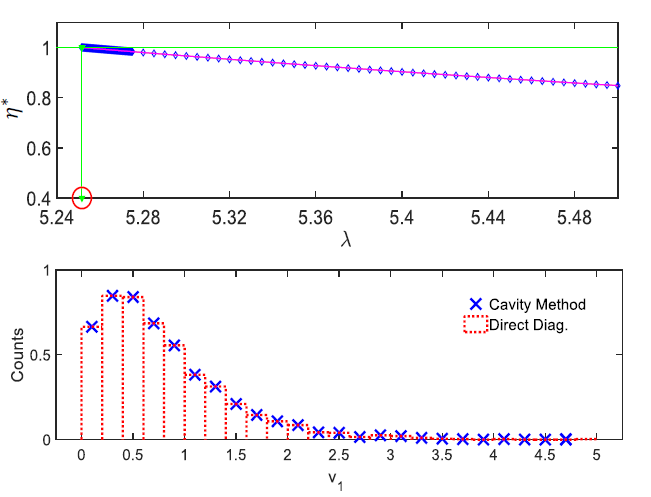}

\par\end{centering}

\begin{centering}
\caption{Cavity single instance. The example refers to a single  Erd\H{o}s-R\'enyi adjacency matrix of size $N=2000$ and mean degree $c=4$.  In the upper panel, the plot of the ratio $\eta^{\star}$ (see \eqref{etastar}) as a function of the parameter $\lambda$: $\lambda$ is lowered (blue diamonds) until $\eta^{\star}=1$. In correspondence of this value, $\lambda=\lambda_1$ (red circle).  The cavity method predicts the value $\lambda_1=5.251599$, to be compared with the value $\lambda_1^{\mathrm{diag}}=5.251575$ obtained by direct diagonalisation, resulting in a relative error of $0.001\%$.
 In the lower panel, the histogram of top eigenvector components of the same matrix as predicted by \eqref{eq:components_single} shows perfect agreement with the components obtained by direct diagonalisation.}
\label{fig:cavity_single_instance}
\par\end{centering}
\centering{}
\end{figure}

Collecting the $H_{j}^{(i)}$'s in a $2M=\sum_{i=1}^{N}k_{i}$ dimensional vector, Eq. \eqref{eq:H} can be rewritten as a linear vector iteration driven by $B$ as

\begin{equation}
H_{j}^{(i)}=\sum_{(k,\ell)}B_{(i,j),(k,\ell)}H_{\ell}^{(k)}\ ,\label{eq:iteration1}
\end{equation}
where the entries $B_{(i,j),(k,\ell)}$ are defined in \eqref{eq:non-back}.
Relabelling with a new, single index $a$ any pair
of connected indices $(i,j)$, \eqref{eq:iteration1} reads 

\begin{equation}
H_{a}=\sum_{b=1}^{2M}B_{ab}H_{b}\,,
\end{equation}
which can interpreted as a vector linear iteration,

\begin{equation}
\underline{H}_{t}=B\underline{H}_{t-1}\ ,\label{eq:sync_update_h}
\end{equation}
with the index $t$ labelling each iteration.

Starting from a certain initial condition $\underline{H}_{0}$, the solution of \eqref{eq:H} is obtained after successive iterations according \eqref{eq:sync_update_h} until $\underline{H}_{t}$ stabilises. The stability can be assessed by looking at the norm of the vector $\underline{H}_{t}$. After a suitable number of iterations $t$, expanding the initial condition vector in the basis $\left\{ \bm{b}_{i}\right\} $ formed by the right eigenvectors of
$B$, the leading contribution is expressed in terms of its top eigenpair 

\begin{equation}
\underline{H}_{t}=B^{t}\underline{H}_{0}=B^{t}\left(\sum_{i=1}^{2M}c_{i}(0)\bm{b}_{i}\right)\approx c_{1}(0)\gamma_{1}^{t}\bm{b}_{1}\,,\label{eq:asympt_h}
\end{equation}
where the contributions coming from the other eigenpairs $\left\{ \bm{b}_{i},\gamma_{i}\right\} $
 are  exponentially suppressed, all the other eigenvalues of $B$ being
smaller than $\gamma_{1}$. 

The ratio $\eta_{t}$ of the norms of two successive iterations approaches a constant value $\eta^{\star}$ as $t\rightarrow\infty$, corresponding to the absolute value of largest eigenvalue of $B$,

\begin{equation}
\eta_{t}=\frac{\left\Vert \underline{H}_{t}\right\Vert }{\left\Vert \underline{H}_{t-1}\right\Vert }=\frac{\left\Vert B\underline{H}_{t-1}\right\Vert }{\Vert\underline{H}_{t-1}\Vert}\to\eta^{\star}=\left|\gamma_{1}\right|\ .\label{etastar}
\end{equation}

Thus, the convergence of \eqref{eq:H} is attained when the value of $\eta^{\star}=\left|\gamma_{1}\right|$ reaches $1$ as $\lambda$ approaches $\lambda_{1}$ from above. We again recall that $\lambda=\lambda_{1}$ is the smallest possible value such that the cavity partition function \eqref{eq:cavity_partition} is well defined, and so the actual value $\lambda_1$ can be found by  asymptotic extrapolation.
Figure \ref{fig:cavity_single_instance} shows an example of this procedure.

We remark that the procedure above holds only if the largest eigenvalue of $B$ is real:
if it is complex, there will be a pair of complex conjugate
first eigenvalues, i.e. those with the largest norm, which dictate the asymptotic behavior of \eqref{eq:sync_update_h}. In
this case, the bi-orthonormal basis of left and right eigenvectors must
be taken into account 
\begin{equation}
\underline{H}_{t}\approx c_{1}(0)\gamma_{1}^{t}\bm{b}_{1}+c_{2}(0)(\gamma_{1}^{\star})^{t}\bm{b}_{1}^{\star}\ ,
\end{equation}
where the coefficients $c_{1}(0)$ and $c_{2}(0)$ are in general
complex. Therefore, the quantity $\eta_{t}$ does not approach 
a steady limit for large $t$ in this case, and oscillations arise. In fact, it
can be shown that

\begin{equation}
\frac{\left\Vert \underline{H}_{t}\right\Vert ^{2}}{\left\Vert \underline{H}_{t-1}\right\Vert ^{2}}=\eta^{2}\frac{\left[|c_{1}|^{2}+|c_{2}|^{2}+2a\cos\left(2\phi t+\psi\right)\right]}{\left[|c_{1}|^{2}+|c_{2}|^{2}+2a\cos\left(2\phi\left(t-1\right)+\psi\right)\right]}\ ,
\end{equation}
where 

\begin{align}
a &= \alpha|c_{1}||c_{2}|\ ,\\
\psi &= \phi_{1}-\phi_{2}+\theta\ .
\end{align}
Here, ($|c_{1}|,|c_{2}|)$ and $(\phi_{1},\phi_{2})$ are the moduli
and phases of the complex coefficients $c_{1}(0)$ and $c_{2}(0)$, $\eta$ is the ratio of the radial part of the vectors $\underline{H}_{t}$ and $\underline{H}_{t-1}$, 
$\alpha$ and $\theta$ are respectively the modulus and the phase
of the dot product between the right (and left) eigenvector $\bm{b}_{1}$
(respectively $\bm{b}_{1}^{\star}$) with itself, and $\rho$ and
$\phi$ are the modulus and phase of the pair of the complex eigenvalues with the largest norm.

In this case, the recursion \eqref{eq:sync_update_h}
does not converge to a single limit, and the cavity formalism does not lead to an acceptable solution.
Therefore, the strongest limitation of the single instance cavity method is that the largest eigenvalue $\gamma_{1}$ of the non-backtracking operator $B$ associated to the matrix $J$ must be real. This restriction unfortunately rules out a variety of interesting sparse matrix ensembles.

\section{\\Exact replica calculation for the largest eigenvalue for any bounded degree distribution $p(k)$.\label{sec:No_shortcuts}}
In this appendix, we show how to get the typical largest eigenvalue with the replica method without any shortcut in the calculation. We will thus employ the distributions \eqref{eq:joint_true} and \eqref{eq:boundedER_connectivity} to perform the averaging w.r.t the matrix ensemble. We recall that the parameter $c$ appearing in \eqref{eq:boundedER_connectivity} stands for the actual mean of the bounded degree distribution of interest, which may in general differ from the parameter $\bar{c}$ of the truncated Poisson distribution (see \eqref{eq:degree_pmf}). They tend to coincide only if $k_{\mathrm{max}}$ is large.
This procedure is general and holds for any graph within the configurational model with degree sequence originated by a finite-mean degree distribution $p(k)$.

Following the same reasoning in Section \ref{sec:replica_eigenvalue}, the replicated partition function is given by \eqref{eq:rep_Z}.
Taking the average w.r.t the joint distribution \eqref{eq:boundedER_connectivity} of matrix entries yields \cite{Kuehn2011}

\begin{align}
\left\langle \exp\left( \frac{\beta}{2}\sum_{a=1}^{n}\sum_{i,j}^{N}v_{ia}J_{ij}v_{ja}\right) \right\rangle _{J}&=\frac{1}{\mathcal{M}}\int_{-\pi}^{\pi}\left( \prod_{i=1}^{N}\frac{\mathrm{d}\phi_i}{2\pi}\right)\exp\left( -\mathrm{i}\sum_i\phi_i k_i\right)\nonumber \\
&\times\exp\left[ \frac{c}{2N}\sum_{i,j}\left(\left\langle \mathrm{e}^{\beta K\sum_{a}v_{ia}v_{ja}+\mathrm{i}(\phi_i+\phi_j)}\right\rangle _{K}-1\right)\right]\ ,\label{eq:B_ensemble_average}
\end{align}
where $\left\langle\cdot \right\rangle _{K}$ denotes averaging over
the single-variable pdf $p\left(K\right)$ characterising
the i.i.d. bond weights $K_{ij}$. A Fourier representation of the Kronecker deltas expressing the degree constraints in \eqref{eq:boundedER_connectivity} has been employed. As in Section \ref{sec:replica_eigenvalue}, we also employ a Fourier representation of the Dirac delta enforcing the normalisation constraint. The replicated partition function thus becomes

\begin{align}
\left\langle Z^{n}\right\rangle _{J} & =\frac{1}{\mathcal{M}}\left(\frac{\beta}{4\pi}\right)^{n}\int\left(\prod_{a=1}^{n}\mathrm{d}\bm{v}_{a}\mathrm{d}\lambda_a\right)\exp\left( \mathrm{i}\frac{\beta}{2}N\sum_{a}\lambda_{a}\right) \exp\left( -\mathrm{i}\frac{\beta}{2}\sum_{a}\sum_{i}\lambda_{a}v_{ia}^{2}\right)\nonumber \\
 &\times \int_{-\pi}^{\pi}\left( \prod_{i=1}^{N}\frac{\mathrm{d}\phi_i}{2\pi}\right)\exp\left( -\mathrm{i}\sum_i\phi_i k_i\right)\exp\left[ \frac{c}{2N}\sum_{i,j}\left(\left\langle \mathrm{e}^{\beta K\sum_{a}v_{ia}v_{ja}+\mathrm{i}(\phi_i+\phi_j)}\right\rangle _{K}-1\right)\right] \ . \label{B_rep_Z}
\end{align}
In order to decouple sites, we introduce the functional order parameter

\begin{equation}
\rho\left(\vec{v},\phi\right)=\frac{1}{N}\sum_{i=1}^{N}\delta\left(\phi-\phi_i\right)\prod_{a=1}^{n}\delta\left(v_{a}-v_{ia}\right)\,,
\end{equation}
where the symbol $\vec{v}$ denotes a $n$-dimensional
vector in replica space. We then consider its integrated version \cite{Kuehn2011}
\begin{equation}
\rho\left(\vec{v}\right)=\int \mathrm{d}\phi~ \mathrm{e}^{\mathrm{i}\phi}\rho\left(\vec{v},\phi\right) =\frac{1}{N}\sum_{i=1}^{N}\mathrm{e}^{\mathrm{i}\phi_i}\prod_{a=1}^{n}\delta\left(v_{a}-v_{ia}\right)\ ,\label{eq:integrated_rho}
\end{equation}
and enforce the latter definition using the integral identity

\begin{equation}
1=\int N\mathcal{D}\rho\mathcal{D}\hat{\rho}\exp\left\{ -\mathrm{i}\int\mathrm{d}\vec{v}\ \hat{\rho}\left(\vec{v}\right)\left[N\rho\left(\vec{v}\right)-\sum_{i}\mathrm{e}^{\mathrm{i}\phi_i}\prod_{a=1}^{n}\delta\left(v_{a}-v_{ia}\right)\right]\right\} \ .
\end{equation}
In terms of the integrated order parameter \eqref{eq:integrated_rho} and its conjugate, the replicated partition function can be written as

\begin{align}
\nonumber &\left\langle Z^{n}\right\rangle _{J}  =\frac{1}{\mathcal{M}}\left(\frac{\beta}{4\pi}\right)^n N\int\mathcal{D}\rho\mathcal{D}\hat{\rho}\mathrm{d}\vec{\lambda}\exp\left( -\mathrm{i}N\int\mathrm{d}\vec{v}\hat{\rho}\left(\vec{v}\right)\rho\left(\vec{v}\right)\right)\exp\left({ \mathrm{i}\frac{\beta}{2}N\sum_{a}\lambda_{a}} \right)\\
\nonumber & \times\exp\left[\frac{Nc}{2}\int\mathrm{d}\vec{v}\mathrm{d}\vec{v^\prime}\rho(\vec{v})\rho(\vec{v^\prime})\left(\left\langle \mathrm{e}^{\beta K\sum_{a}v_{a}v_{a}^{'}}\right\rangle _{K}-1\right)\right] 
\int_{-\pi}^{\pi}\left( \prod_{i=1}^{N}\frac{\mathrm{d}\phi_i}{2\pi}\right)\mathrm{e}^{ -\mathrm{i}\sum_i\phi_i k_i}\\
 &\int\prod_{a=1}^{n}\mathrm{d}\bm{v}_{a}\exp\left[ -\mathrm{i}\frac{\beta}{2}\sum_{a}\sum_{i}\lambda_{a}v_{ia}^{2}+ \mathrm{i}\sum_{i}\mathrm{e}^{\mathrm{i}\phi_i}\int\mathrm{d}\vec{v}\hat{\rho}\left(\vec{v}\right)\prod_{a=1}^{n}\delta\left(v_{a}-v_{ia}\right)\right] \ .
\end{align}
The multiple integral in the last line above is the product of $N$ $n$-dimensional integrals, each related to a degree $k_i$. It can be written as

\begin{align}
I=&\prod_{i=1}^{N}\int_{-\pi}^{\pi}\frac{\mathrm{d}\phi_i}{2\pi}\int\mathrm{d}\vec{v}_i\exp\left( -\mathrm{i}\phi_i k_i-\mathrm{i}\frac{\beta}{2}\sum_{a}\lambda_{a}v_{ia}^{2}+\mathrm{i}\hat{\rho}(\vec{v}_i)\mathrm{e}^{\mathrm{i}\phi_i}\right)\nonumber\\
=&\exp\left[ \sum_{i=1}^N \mathrm{Log} \int\mathrm{d}\vec{v}_i \exp\left(-\mathrm{i}\frac{\beta}{2}\sum_{a}\lambda_{a}v_{ia}^{2} \right)I[k_i,\vec{v}_i]\right]\ ,\label{eq:single_site_1}
\end{align}
where $\mathrm{Log}$ denotes the principal branch of the complex logarithm, and
\begin{equation}
I[k_i,\vec{v}_i]=\int_{-\pi}^{\pi}\frac{\mathrm{d}\phi_i}{2\pi}\exp\left(  -\mathrm{i}\phi_i k_i+\mathrm{i}\hat{\rho}(\vec{v}_i)\mathrm{e}^{\mathrm{i}\phi_i}\right)\ .
\end{equation}
Each of the $\phi_i$ integrals can be performed by rewriting the last exponential factor as a power series, viz.
\begin{align}
I[k_i,\vec{v}_i]=&\int_{-\pi}^{\pi}\frac{\mathrm{d}\phi_i}{2\pi}\mathrm{e}^{-\mathrm{i}\phi_i k_i}\sum_{s=0}^{\infty} \frac{\left(\mathrm{i}\hat\rho(\vec{v}_i)^s\right)}{s!}\mathrm{e}^{\mathrm{i}s\phi_i}
=\sum_{s=0}^{\infty} \frac{\left(\mathrm{i}\hat\rho(\vec{v}_i)^s\right)}{s!}\delta_{s,k_i}
=\frac{\left(\mathrm{i}\hat\rho(\vec{v}_i)^{k_i}\right)}{k_i!}\;\;\;\;\; \forall k_i\ ,
\end{align}
with $i=1,\ldots,N$.
Thus, by invoking the Law of Large Numbers, the single site integral $I$ \eqref{eq:single_site_1} can be expressed as
\begin{align}
I=&\exp\left[ \sum_{i=1}^N \mathrm{Log} \int\mathrm{d}\vec{v}_i \exp\left(-\mathrm{i}\frac{\beta}{2}\sum_{a}\lambda_{a}v_{ia}^{2} \right)\frac{\left(\mathrm{i}\hat\rho(\vec{v}_i)^{k_i}\right)}{k_i!}\right]\nonumber\\
=&\exp N \sum_{k=k_{\mathrm{min}}}^{k_{\mathrm{max}}} p(k)\left[\mathrm{Log}\int\mathrm{d}\vec{v}\exp\left(-\mathrm{i}\frac{\beta}{2}\sum_a\lambda_a v_a^2\right)(\mathrm{i}\hat\rho(\vec{v}))^k-\mathrm{Log}(k!)\right]\ ,\label{eq:single_site_2}
\end{align}
where we have used 
\begin{equation}
\frac{1}{N}\sum_{i=1}^{N}\mathrm{Log}f(k_i)\simeq\sum_{k=k_{\mathrm{min}}}^{k_{\mathrm{max}}} p(k) \mathrm{Log}f(k)\ ,
\end{equation}
where $p(k)$ is the actual degree distribution of the graph. 

As in Section \ref{sec:replica_eigenvalue}, the replicated partition function takes a form amenable to a saddle point evaluation for large $N$ 

\begin{equation}
\left\langle Z^{n}\right\rangle _{J}\propto\int\mathcal{D}\rho\mathcal{D}\hat{\rho}\mathrm{d}\vec{\lambda}\exp\left(NS_{n}[\rho,\hat{\rho},\vec{\lambda}]\right)\ ,
\end{equation}
where
\begin{equation}
S_{n}[\rho,\hat{\rho},\vec{\lambda}]=S_{1}\left[\rho,\hat{\rho}\right]+S_{2}\left[\rho\right]+S_{3}(\vec{\lambda})+S_{4}[\hat{\rho},\vec{\lambda}]\ .\label{eq: action}
\end{equation}
The terms $S_1, S_2$ and $S_3$ are equal to those found in Section \ref{sec:replica_eigenvalue}, respectively \eqref{eq:S1}, \eqref{eq:S2} and \eqref{eq:S3}, whereas
\begin{equation}
S_{4}[\hat{\rho},\vec{\lambda}] =\sum_{k=k_{\mathrm{min}}}^{k_{\mathrm{max}}} p(k)\left[\mathrm{Log}\int\mathrm{d}\vec{v}\exp\left(-\mathrm{i}\frac{\beta}{2}\sum_a\lambda_a v_a^2\right)(\mathrm{i}\hat\rho(\vec{v}))^k-\mathrm{Log}(k!)\right] \ .
\end{equation}

As in Section \ref{sec:replica_eigenvalue}, we then search for replica-symmetric saddle-point solutions written in the form of superpositions of uncountably infinite Gaussians with a non-zero mean, 

 \begin{align}
\lambda_{\bar{a}} &=\lambda\qquad\forall \bar{a}=1,\ldots,n\ ,\\
\rho^\star(\vec{v}) &=\rho_0\int\mathrm{d}\omega\mathrm{d}h\ \pi\left(\omega,h\right)\prod_{a=1}^{n}\frac{1}{Z_\beta(\omega,h)}\exp\left[ -\frac{\beta}{2}\omega v_{a}^{2}+\beta hv_{a}\right] \ ,\label{eq:B_ansatz_rep_1}\\
\hat{\rho}^\star(\vec{v}) &=\hat{\rho}_0\int\mathrm{d}\hat{\omega}\mathrm{d}\hat{h}\ \hat{\pi}(\hat{\omega},\hat{h})\prod_{a=1}^{n}\exp\left[ \frac{\beta}{2}\hat{\omega}v_{a}^{2}+\beta\hat{h}v_{a}\right] \ ,\label{eq:B_ansatz_rep_2}
\end{align}
where 

\begin{equation}
Z_\beta(x,y)=\sqrt{\frac{2\pi}{\beta x}}\exp\left(\frac{\beta y^{2}}{2 x}\right)\ ,\label{eq:zomega_erB}
\end{equation}
and -- with a modest amount of foresight -- we use the same notation as before for the distributions $\pi$ and $\hat\pi$.
The $\rho_0$ and $\hat\rho_0$ are determined such that the distributions $\pi(\omega,h)$ and $\hat\pi(\hat\omega,\hat{h})$ are normalised. The $\rho_0$ in \eqref{eq:B_ansatz_rep_1} is needed since $\rho^\star(\vec{v})$ is the saddle-point expression of the integrated order parameter.

Rewriting the action in terms of $\pi$ and $\hat\pi$, after performing the $\vec{v}$-integrations, and extracting the leading $n\to 0$ contribution yields

\begin{equation}
S_n=S_{1}[\pi,\hat{\pi}]+S_{2}[\pi]+S_{3}(\lambda)+S_{4}[\hat{\pi},\lambda]\ ,\label{eq:action_pihatpiB}
\end{equation}

with 
\begin{align}
S_{1}[\pi,\hat{\pi}] & =-\mathrm{i}\rho_0\hat\rho_0-\mathrm{i}\rho_0\hat\rho_0 n\int\mathrm{d}\pi(\omega,h)\mathrm{d}\hat{\pi}(\hat{\omega},\hat{h})\ln\frac{Z_\beta(\omega-\hat{\omega},h+\hat{h})}{Z_\beta(\omega,h)}\ ,\label{eq:B_S1_pi} & {}\\
S_{2}[\pi] & =\frac{c}{2}\left(\rho_0^2-1 \right)+n\frac{c}{2}\rho_0^2 \int\mathrm{d}\pi(\omega,h)\mathrm{d}\pi(\omega',h')\left\langle \ln\frac{Z^{(2)}_\beta\left(\omega,\omega',h,h',K\right)}{Z_\beta\left(\omega,h\right)Z_\beta\left(\omega',h'\right)}\right\rangle _{K}\ ,\label{eq:B_S2_pi}\\
S_{3}(\lambda) & =\mathrm{i}\frac{\beta}{2}n\lambda\ ,\label{eq:B_S3_pi}\\
S_{4}[\hat{\pi},\lambda] & =c\mathrm{Log}(\mathrm{i}\hat\rho_0)-\sum_{k=0}^{k_{\mathrm{max}}} p(k)\mathrm{Log}(k!)+n\sum_{k=0}^{k_{\mathrm{max}}} p(k)\int\{ \mathrm{d}\hat{\pi}\} _{k}~\mathrm{Log}~Z_\beta\left(\mathrm{i}\lambda-\{ \hat{\omega}\} _{k},\{ \hat{h}\} _{k}\right)\label{eq:B_S4_pi}\ ,
\end{align}
where we have taken into account that $k_{\mathrm{min}}=0$ and we have introduced the shorthands 
\begin{equation}
Z^{(2)}_\beta(\omega,\omega',h,h',K)=Z_\beta(\omega',h')Z_\beta\left(\omega-\frac{K^{2}}{\omega'},h+\frac{h'K}{\omega'}\right)
\end{equation}
and $\{\mathrm{d}\hat{\pi}\} _{s}=\prod_{\ell=1}^{s}\mathrm{d}\hat{\omega}_{\ell}\mathrm{d}\hat{h}_{\ell}\hat{\pi}(\hat{\omega}_{\ell},\hat{h}_{\ell})$,
along with $\{ \hat{\omega}\} _{s}=\sum_{\ell=1}^{s}\hat{\omega}_{\ell}$
and $\{ \hat{h}\} _{s}=\sum_{\ell=1}^{s}\hat{h}_{\ell}$.

We note that the action contains $\mathcal{O}(1)$ and $\mathcal{O}(n)$ terms as $n\to 0$: the $\mathcal{O}(1)$ terms are cancelled by the $\mathcal{O}(1)$ terms arising from the evaluation of the normalisation constant $\mathcal{M}$ at the saddle-point.
Indeed, by following a very similar reasoning as in \eqref{eq:B_ensemble_average}, we find that
\begin{equation}
\mathcal{M}=\int_{-\pi}^{\pi} \left( \prod_{i=1}^{N} \frac{\mathrm{d}\phi_i}{2\pi} \right) \mathrm{e}^{-\mathrm{i}\sum_i \phi_i k_i} \exp\left[ \frac{c}{2N} \sum_{i,j}\left( \mathrm{e}^{\mathrm{i}(\phi_i+\phi_j)}-1\right)\right]\ .\label{eq:M_1}
\end{equation}

We then introduce in \eqref{eq:M_1} the scalar order parameter
\begin{equation}
\rho_0=\frac{1}{N}\sum_{i=1}^N\mathrm{e}^{\mathrm{i}\phi_i}
\end{equation}
via the integral representation
\begin{equation}
1=\int N\frac{ \mathrm{d}\rho_0 \mathrm{d} \hat\rho_0}{2\pi}\exp\left[-\mathrm{i}\hat\rho_0\left(N\rho_0-\sum_i \mathrm{e}^{\mathrm{i}\phi_i}\right) \right] \ .
\end{equation}

By using the same argument as in \eqref{eq:single_site_2}, the normalisation constant $\mathcal{M}$ can be written in a form amenable to a saddle-point evaluation,

\begin{align}
\mathcal{M}=&\int N\frac{ \mathrm{d}\rho_0 \mathrm{d} \hat\rho_0}{2\pi}\exp\left[N\left(-\mathrm{i}\rho_0\hat\rho_0+\frac{c}{2}(\rho_0^2-1)+c\mathrm{Log}(\mathrm{i}\hat\rho_0)-\sum_{k=0}^{k_{\mathrm{max}}}p(k)\mathrm{Log}(k!) \right) \right]\nonumber\\
=&\int N\frac{ \mathrm{d}\rho_0 \mathrm{d} \hat\rho_0}{2\pi}\exp\left[N S_{\mathcal{M}}(\rho_0,\hat\rho_0)\right]\ . \label{eq:M_sp}
\end{align}
The stationarity conditions for $S_{\mathcal{M}}$ are
\begin{equation}
\frac{\partial S_{\mathcal{M}}}{\partial \rho_0}=0 \Rightarrow \mathrm{i}\hat\rho_0=c\rho_0\ ,\label{eq: M_sp_1}
\end{equation}
and
\begin{equation}
\frac{\partial S_{\mathcal{M}}}{\partial \hat\rho_0}=0 \Rightarrow \mathrm{i}\rho_0=\frac{c}{\hat\rho_0}\ . \label{eq: M_sp_2}
\end{equation}
entailing that 
\begin{align}
\mathrm{i}\rho_0\hat\rho_0=&c\ , \label{eq:M_sp_11}\\ 
\rho_0^2=1\label{eq:M_sp_22} \ .
\end{align}
The  two conditions above exhibit a gauge invariance\cite{Kuehn2011}. Once the same gauge has been chosen for the saddle-point solution of $\mathcal{M}$ and the $\mathcal{O}(1)$ terms of the action  \eqref{eq:action_pihatpi} in the numerator, they cancel out so that the action \eqref{eq:action_pihatpi} is $\mathcal{O}(n)$ as expected.

Thus, taking into account the cancellation coming from \eqref{eq:M_sp_11} and \eqref{eq:M_sp_22}, the action terms in  \eqref{eq:action_pihatpi} read exactly as those found in Section \ref{sec:replica_eigenvalue}, thus proving that the ``shortcut" derivation in \ref{sec:replica_eigenvalue} is perfectly legitimate. According to the present derivation, the degree distribution $p(k)$ appearing in the single-site term $S_4$ is already the true degree distribution of the graph, and does not require any a posteriori correction.

\section*{\textemdash \textemdash \textemdash \textemdash \textemdash \textendash{}}



\end{document}